# Stabilization and high thermoelectric performance of high-entropy-type cubic AgBi(S, Se, Te)$_2$

Asato Seshita[a,†], Aichi Yamashita[a,b,†,*], Takeshi Fujita[b,c], Takayoshi Katase[d], Akira Miura[e], Yuki Nakahira[f], Chikako Moriyoshi[g], Yoshihiro Kuroiwa[g], Yoshikazu Mizuguchi[a]

[a] *Department of Physics, Tokyo Metropolitan University, 1-1 Minami-Osawa, Hachioji-shi, Tokyo 192-0379 Japan*
[b] *School of Engineering Science, Kochi University of Technology, Kami, Kochi 782-8502, Japan*
[c] *Center for Nanotechnology, Kochi University of Technology, Kami, Kochi 782-8502, Japan*
[d] *MDX Research Center for Element Strategy, International Research Frontiers Initiative, Tokyo Institute of Technology, 4259 Nagatsuta, Midori, Yokohama 226-8501, Japan*
[e] *Faculty of Engineering, Hokkaido University, Kita-13, Nishi-8, Kita-ku, Sapporo, Hokkaido 060-8628, Japan*
[f] *Synchrotron Radiation Research Center, National Institutes for Quantum Science and Technology (QST), 1-1-1 Koto, Sayo-cho, Sayo-gun, Hyogo 679-5148, Japan*
[g] *Graduate School of Advanced Science and Engineering, Hiroshima University, 1-3-1 Kagamiyama, Higashihiroshima, Hiroshima 739-8526, Japan*

**Abstract**

As thermoelectric generators can convert waste heat into electricity, they play an important role in energy harvesting. The metal chalcogenide AgBiSe$_2$ is one of the high-performance thermoelectric materials with low lattice thermal conductivity ($\kappa_{\text{lat}}$), but it exhibits temperature-dependent crystal structural transitions from hexagonal to rhombohedral, and finally a cubic phase as the temperature rises. The high figure-of-merit $ZT$ is obtained only for the high-temperature cubic phase. In this study, we utilized the high-entropy-alloy (HEA) concept for AgBiSe$_2$ to stabilize the cubic phase throughout the entire temperature range with enhanced thermoelectric performance. We synthesized high-entropy-type AgBiSe$_{2-2x}$S$_x$Te$_x$ bulk polycrystals and realized the stabilization of the cubic phase from room temperature to 800 K for $x \geq 0.6$. The ultra-low $\kappa_{\text{lat}}$ of 0.30 Wm$^{-1}$K$^{-1}$ and the high peak $ZT \sim 0.9$ at around 750 K were realized for cubic AgBiSe$_{2-2x}$S$_x$Te$_x$ without carrier tuning. In addition, the average $ZT$ value of $x = 0.6$ and 0.7 for the temperature range of 360–750 K increased to 0.38 and 0.40, respectively, which are comparable to the highest previously reported values.

[†]These two authors contributed equally to this work
*Corresponding author: Aichi Yamashita
E-mail: aichi@tmu.ac.jp

**Introduction**

Thermoelectric technology can directly convert waste heat into electrical energy and vice versa. The core of this technology relies on thermoelectric materials, and the performance is evaluated by a dimensionless figure-of-merit ($ZT$), expressed as $ZT = S^2T/\rho\kappa_{tot}$. Here, $S$, $\rho$, $T$, and $\kappa_{tot}$ represent the Seebeck coefficient, electrical resistivity, absolute temperature, and total thermal conductivity, respectively. $\kappa_{tot}$ encompasses the lattice ($\kappa_{lat}$) and electronic ($\kappa_{ele}$) contributions [1]. Therefore, a high performance can be achieved for a thermoelectric material by a simultaneous decrease in $\rho$ and $\kappa_{tot}$, which may seem contradictory. To overcome this contradiction, several strategies have been proposed for achieving a selective suppression of $\kappa_{lat}$ with minimal alteration of $\rho$, such as employing a multiscale hierarchical architecture to provide full-spectrum phonon scattering [2] or phonon scattering through nanostructure engineering [3].

The AgBiSe$_2$ metal-chalcogenide has attracted much attention as a promising candidate for a high-performance thermoelectric material due to its intrinsically low $\kappa_{lat}$ of ~ 1 Wm$^{-1}$K$^{-1}$ at room temperature [4–12]. AgBiSe$_2$ exhibits a structural phase transition from hexagonal to rhombohedral, followed by a transition to the cubic phase, as the temperature increases (see Fig. 1). An excellent TE performance of $ZT \sim 1$ at 773 K has been reported for the cubic phase with Nb doping [6], which makes AgBiSe$_2$ a promising n-type candidate for medium-temperature thermoelectric applications.

However, these complicated structural transitions are a drawback for module applications. There is a concern that the thermoelectric performance and mechanical strength will deteriorate if the crystal structure differs between the low- and high-temperature regions, owing to the temperature difference during power generation. To date, it has been reported that the partial substitution of Se sites with its isovalent elements S or Te reduces the structural phase transition temperature to the cubic phase [9,13], but stabilization of the cubic structure has not yet achieved. Although the cubic structure can be stabilized by substituting 30% of the Bi site with the homologous element Sb even at room temperature [14], the thermoelectric performance decrease because of the high electrical resistivity, possibly due to an increase in the band gap energy due to Sb substitution.

A novel concept for stabilization by mixing entropy has been proposed, known as high-entropy-alloy (HEAs) [15,16]. HEAs are typically defined as alloys containing at least five elements with concentrations between 5 and 35 at%, resulting in a high mixing entropy ($\Delta S_{mix}$) defined as $\Delta S_{mix} = -R\Sigma_i c_i \ln c_i$, where $c_i$ and $R$ are the compositional ratio and the gas constant, respectively. It has been reported that in HEA, the Gibbs free energy decreases and the structural stability improves with increasing $\Delta S_{mix}$. HEAs have attracted considerable attention in the fields of materials science and engineering due to their excellent mechanical performance under extreme conditions [15, 16]. While this concept was originally developed for "alloys", we have extended the concept of HEA to other compounds as high-entropy-type (HE-type) compounds, for instance, layered compounds and NaCl-type metal chalcogenide, as superconductors with high $\Delta S_{mix}$ [17–25]. Furthermore, as an efficient way to increase the total $\Delta S_{mix}$, we proposed multi-site alloying of compounds and its evaluation by summing the $\Delta S_{mix}$ of each alloying site [23]. A new NaCl-type MCh (Ch: S, Se, Te) thermoelectric material with the highest $\Delta S_{mix}$ of $2.0R$, which exceeds the typical $\Delta S_{mix}$ of 6 equimolar elements for a single site, was achieved by this method [17]. In the same period, n-type MCh of $Pb_{0.99-y}Sb_{0.012}Sn_ySe_{1-2x}Te_xS_x$ was reported with a high $ZT$ value of 1.8 at 900 K [26], and, very recently, an extremely high $ZT$ value of 2.7 for HE-type GeTe [27] was also reported from the same research group. In the aforementioned papers, all-scale scattering sources for heat-carrying phonons or localized phonons from the entropy-induced disorder dampening the propagation of transverse phonons were proposed as the origin of the large reduction in $\kappa_{lat}$. In 2023, cubic phase stabilization for $AgBiSe_2$ was also reported for SnTe alloying [28], which resulted in a $\Delta S_{mix}$ of $1.4R$, which is categorized as a middle-entropy material. However, the bipolar effect, which is caused by alloying with a very narrow band gap material such as SnTe, remains a challenging issue for achieving high thermoelectric performance in the high-temperature region.

It has been clearly shown that the incorporation of HEAs into thermoelectric materials can be an effective strategy for achieving both the stabilization of the crystal structure and a high performance [27]. In this study, we synthesized high-entropy-type (HE-type) $AgBiSe_2$ compounds that incorporate the concept of HEAs to stabilize the high-temperature cubic structure by anionic substitution of the Se site with S and Te. the temperature dependencies of the crystal structure revealed the stabilization of the cubic phase for $x \geq 0.6$ from room temperature to 800 K. In addition to the structural stabilization, our study presents the conclusion that a high thermoelectric performance can be realized by the synergy of an ultra-low $\kappa_{lat}$ due to effective point defect scattering and good transport properties with a suppression of the bipolar effect utilizing the simultaneous substitution of the Ch site.

## Methods

### Synthesis

Polycrystalline samples of $AgBiSe_{2-2x}S_xTe_x$ with $x$ = 0, 0.1, 0.2, 0.3, 0.4, 0.5, 0.6, 0.7, 0.8 were synthesized by solid-state reactions using elemental Ag powders (99.9%, Kojundo Kagaku) and grains of Bi (99.999%, Kojundo Kagaku), Se (99.999%, Kojundo Kagaku), S (99.999%, Kojundo Kagaku), and Te (99.9995%, Kojundo Kagaku). A stoichiometric ratio of these materials was mixed, pelletized, and then heated at 500°C for 15 h in an evacuated quartz tube twice. To obtain high-density samples, hot pressing was performed at 500°C for 30 min under a uniaxial pressure of 60 MPa. The relative density of the samples was beyond 95%.

### Sample assessment

The actual compositions of the samples were estimated using energy-dispersive X-ray spectroscopy (EDX) on a TM-3030 (Hitachi Hightech), which is a scanning electron microscope (SEM) equipped with a Swift-ED analyzer (Oxford). The phase purity and crystal structure of the samples at room temperature were examined by means of synchrotron powder X-ray diffraction (SPXRD) at the BL13XU (Proposal No. 2023A1042) beamline of SPring-8. The SPXRD experiments were performed at room temperature with a sample rotator system, and the diffraction data were collected using a high-resolution LAMBDA 750 K detectors [29] with a step of $2\theta = 0.006°$. The crystal structure parameters were refined using the Rietveld method with the RIETAN-FP software [30], and the crystal structure was visualized using VESTA software [31].

**Phase transition temperature**

A temperature-composition phase diagram of AgBiSe$_{2-2x}$S$_x$Te$_x$ was determined using SPXRD at the BL02B2 (Proposal No. 2022B105) beamline of the SPring-8. The SPXRD experiments were performed from 353 K to 873 K, and the wavelength of the beam was determined to be $\lambda$ =0.4962908(3) Å using a CeO$_2$ standard. The phase determination of each temperature was carried out by comparison of the relative intensity and peak positions of (110), (018), and (220) peaks. At the hexagonal-rhombohedral phase transition boundary, the peak positions of (110) and (018) peaks were distant in the case of hexagonal phase, while the rhombohedral phase had a narrow peak position between (110) and (018) peaks. At the rhombohedral-cubic phase transition boundary, only the cubic phase had (220) between (110) and (018) peaks, and then (110) and (018) peaks gradually disappeared.

**Transmission electron microscopy**

The microstructures were characterized using transmission electron microscopy (TEM) with JEM-ARM200F NEOARM (JEOL), which is equipped with aberration correctors for the image- and probe-forming lens systems (CEOS GmbH), and energy dispersive X-ray spectroscopy (EDS) with JED-2300T (JOEL). The TEM and scanning TEM (STEM) experiments were conducted at an accelerating voltage 200 kV.

**Electrical transport properties**

The $\rho$ and $S$ were simultaneously measured using the four-probe method with a ZEM-3 instrument (Advance Riko) from room temperature to approximately 500°C in a He atmosphere. To make the structural transition complete, measurements were kept for 1 hour near the structural transition temperature to cubic phase.

**Thermal transport properties**

The $\kappa_{tot}$ was calculated using $\kappa_{tot} = DC_p d_s$, where $D$ is the thermal diffusivity, $C_p$ is the specific heat, and $d_s$ is the sample density, respectively. $D$ was measured by means of the laser-flash method with a TC1200-RH instrument (Advance Riko). The value of $C_p$ was estimated using the Dulong-Petit model, $C_p = 3nR$, where $n$ is the number of atoms per formula unit.

**Hall coefficient**

The room temperature Hall coefficient ($R_H$) was examined by the formula; $\Delta\rho_{xy} = R_H B$ ($\rho_{xy}$: Hall resistivity, $B$: magnetic flux density). To remove the effective longitudinal resistivity, the $\rho_{xy}$ was estimated by the formula; $\Delta\rho_{xy}(B) = (\rho_{xy}(B) - \rho_{xy}(-B))/2$, and it was measured using four probe method with physical property measurement system (PPMS, Quantum Design) under magnetic fields up to 5 T.

**Results and discussion**

**Chemical composition and Structural characterization**

Room temperature SPXRD patterns of AgBiSe$_{2-2x}$S$_x$Te$_x$ with $x = 0$–0.8 are shown in Fig. 2. All peaks of the $x = 0$–0.5 samples were indexed as hexagonal structure (space group: $P\bar{3}m1$, #164), which is the typical room-temperature crystal structure of AgBiSe$_2$ phase. On the other hand, those of $x = 0.6$–0.8 were indexed as cubic structure (space group: $Fm\bar{3}m$, #225), which is well known as the high-temperature phase of AgBiSe$_2$. Thus, we successfully obtained the cubic structure of AgBiSe$_{2-2x}$S$_x$Te$_x$. Although $x = 0.8$ showed small amount of Bi$_2$(S, Se, Te)$_3$ impurity phase with 2.7 wt%, no peak split was observed in all samples, indicating the homogeneity of the samples. Successful stabilization of the cubic structure at room temperature would be possibly due to the increase in $\Delta S_{mix}$ in the system [26,27]. The systematic increase of lattice parameter $c$ for hexagonal and $a$ for both hexagonal and cubic with increasing $x$ also indicates that Se atoms were substituted by S and Te atoms (See supplemental Fig. S1). The sudden decrease of lattice parameter $a$ and increase of $c$ for $x = 0.5$ sample would be attributed to the emergency of cubic phase, which may result in the transfer of larger atom from hexagonal to cubic. Note that, multi-phase Rietveld refinement revealed the coexistence of rhombohedral phase with 10% in $x = 0$–0.4 samples, which is comparable to previous reports [8–10], and also 16% of cubic phase and 34% of hexagonal phase were also found in the $x = 0.5$ and 0.6 samples (See supplemental

Fig. S2). Through the Rietveld refinement, anti-site disorder (ASD) [8,9] was also found in the hexagonal structure between Ag 2$d$ and Bi 2$d$ on their respective sites. Fractions of ASD were estimated as around 5% for $x = 0$–0.4 and increased to 11% for $x = 0.5$ samples (See supplemental Fig. S3). The detailed results of Rietveld refinement were summarized in supplemental Fig. S4. The SEM-EDX analyses also revealed that there is no compositional segregation (See Supplemental Figs. S5(a–i)). Note that tiny deviation from the nominal composition, which is indicated by dashed line, was observed for the S element. This chalcogen deficiency would occur during the sintering due to its high volatility. Although very small deviation of S was observed, the actual composition is in good agreement with the nominal composition (see Figs. S5(j–m) and also see Table S1). The estimated $\Delta S_{mix}$ through the actual composition was plotted with both of cubic and hexagonal crystal structure cases (Fig.3(o)). In addition, the $\Delta S_{mix}$ estimated with and without ASD situation were also plotted. The $\Delta S_{mix}$ value for the cubic phase exceeded $1.5R$ for $x \geq 0.3$, and it reached $1.7R$ at $x = 0.7$.

In order to reveal the structural phase transition in the present samples, the temperature-dependence of SPXRD was performed from room temperature to 873 K. Figures 4(a–f) show the temperature dependence of SPXRD patterns around (110) and (018) reflections of the hexagonal structure and (220) reflection of the cubic structure with $x = 0$–0.5. Color plots were made by the intensity of SPXRD patterns (Figs. 4 g–l). From these results, the $T$-$x$ phase diagram of AgBiSe$_{2-2x}$S$_x$Te$_x$ with $x = 0$–0.8 was made (Fig. 5). Here, $T_1$ and $T_2$ denote the structural transition temperatures from hexagonal to rhombohedral and rhombohedral to cubic, respectively. We can clearly see a drastic reduction of $T_1$ with increasing $x$. Note that these $T_1$ and $T_2$ were determined from the temperature at which either rhombohedral or cubic phase became dominant. From the phase diagram and the color plot, it is clear that $x = 0.5$ is the boundary composition between the hexagonal and cubic phases. For the scenario of stabilization of the cubic phase, we deduce that the increase of $\Delta S_{mix}$, so called HE effect, would decreases the Gibbs free energy and stabilize the high-temperature phase of cubic structure. However, detailed study on the HE effects on the Gibbs free energy, which needs the calculation of enthalpy contribution, are needed. In addition, investigation from the aspects of kinetic barrier and atomic diffusion should be required in the future works, and that is beyond the scope of present study.

**Thermoelectric properties**

Figures 6 (a-d) show the temperature dependence of $\rho$, $S$, power factor (PF), $\kappa_{tot}$, $\kappa_{lat}$, and $ZT$ for AgBiSe$_{2-2x}$S$_x$Te$_x$ with $x = 0$–0.8. Here, the $\rho$, $S$, PF, $\kappa_{lat}$, and $\kappa_{lat}$ were plotted separately in upper and lower panels according to the majority phase, hexagonal or cubic, at room temperature. Figure 6(a) shows the temperature dependence of $\rho$. At room temperature, the $\rho$ value of pristine AgBiSe$_2$ was 6.1 m$\Omega$cm, and it remained almost flat until around 500 K (Lower side of Fig. 6 a). An increase tendency of $\rho$ from around 500 K is consistent with the $T_1$ temperature, at which the structural transition from hexagonal to rhombohedral phase occurs. Note that, to elucidate the intrinsic properties of cubic phase, the measurement was kept for 1 hour at 30 K higher than the $T_2$ temperature for $x = 0$–0.5 to make the phase transition complete. Coincidence of all data points with each other within the error indicates the completion of phase transition at this temperature. The value of $\rho$ increased to 33.4 m$\Omega$cm at around 600 K. The observation of drastic increase of $\rho$ around this temperature is also consistent with previous reports [6,7,9,10], and it would be related to the transient region of the phase transition from rhombohedral to cubic. After that, the value of $\rho$ exhibited a reduction tendency and decreased to 19.1 m$\Omega$cm at 748 K. Although the increase of $\rho$ before $T_2$ was relatively smaller than that of $x = 0.0$, similar trend was observed for $x = 0.1$–0.4 according to the $T_1$ and $T_2$ structural phase transition. At room temperature, $\rho$ slightly increased with $x$ amount, and it suddenly increased at $x = 0.5$. In the case of the $x = 0.5$ sample, $\rho$ behaved similar to that of the $x = 0.6$ sample, although it showed a small hump possibly due to a phase transition around 400–500 K. The fraction of ASD was larger than the other hexagonal samples, and the room temperature $\rho$ increased with the disorder. This is why the behavior is quite different from other hexagonal phase samples. On the other hand, the cubic-phase samples behaved like typical semiconductors, even though the $x = 0.6$–0.8 samples were affected by the emergence of the Bi$_2$Ch$_3$ (Ch: S, Se, Te) impurity phase at intermediate temperatures for $x = 0.6$ and 0.7 or room temperature for $x = 0.8$ (see Supplement Fig. S6). The $\rho$ value of $x = 0.6$ was 49.2 m$\Omega$cm at room temperature, and it decreased to 7.7 mΩcm at 748 K with increasing temperature.

The behavior of the $x = 0.7$ and 0.8 samples are discussed with consideration of the transport property of $Bi_2(S,Se,Te)_3$ impurity phase. The behavior of $\rho$, $S$, $\kappa_{tot}$, and $\kappa_{lat}$ of $AgBiSe_{0.4}S_{0.8}Te_{0.8}$ would be explained by classifying in 5 phases as shown in Supplemental Fig. S7. At phase A, all transport properties showed typical semiconductor behavior by $AgBiSe_{0.4}S_{0.8}Te_{0.8}$ and $Bi_2(S, Se, Te)_3$. The $\rho$ increased, and both $\kappa_{tot}$ and $\kappa_{lat}$ decreased, but $S$ remained flat at phase B, thus, carrier concentration should be unchanged, then the boundary scattering and the point defect scattering should be enhanced with increasing amount of the $Bi_2(S, Se, Te)_3$ phase. In the case of phase C, $\rho$ and the absolute value of $S$ decreased, and $\kappa_{tot}$ increased while $\kappa_{lat}$ remained almost flat. The bipolar effect of $Bi_2Te_{2.7-x}S_{0.3}Se_x$ with $x = 0$, 0.2 and 0.4 was observed on the temperature region of phase C; therefore, carrier concentration should be increased by the bipolar effect of $Bi_2(S, Se, Te)_3$ at phase C. Low $\rho$ and high $\kappa_{tot}$ were observed at phase D, indicating that the thermoelectric properties of $Bi_2(S, Se, Te)_3$ was dominant at the phase. At phase E, $\rho$ and the absolute value of $S$ increased, and $\kappa_{tot}$ gradually decreased, thus, the amount of $Bi_2(S, Se, Te)_3$ impurity phase should decrease. As discussed above, the cause of the complicated behavior of $AgBiSe_{0.4}S_{0.8}Te_{0.8}$ was explained by the affection by the $Bi_2(S, Se, Te)_3$ impurity phase.

To further understand the effect of S and Te simultaneous substitution on the electrical properties, Hall measurement was conducted. The Hall coefficient ($R_H$) at room temperature was negative for all samples, indicating n-type conduction. The Hall carrier concentration ($n_H$) was estimated using the formula; $n_H = 1/(eR_H)$, where $e$ is the electronic charge, and the Hall mobility ($\mu_H$) was examined with the formula; $\rho = 1/(en_H \mu_H)$. The obtained values of $n_H$ and $\mu_H$ at room temperature were summarized in Table S2. The $n_H$ and $\mu_H$ have almost inversely proportional relation toward $x$ amount both the hexagonal phase and the cubic phase. The $n_H$ for the pristine $AgBiSe_2$ sample was estimated as $2.2 \times 10^{19}$ cm$^{-3}$, and it increased to $3.4 \times 10^{19}$ cm$^{-3}$ for the $x = 0.1$ sample. It then showed decreasing behavior with the increase of $x$ and reached to $1.2 \times 10^{19}$ cm$^{-3}$ for the $x = 0.5$ sample. This downturn behavior of $n_H$ is consistent with the upturn tendency of $\rho$ at room temperature (Figure S8), indicating the tiny effect of electron carrier concentration possibly due to slight deficiency of Ch site for instance S element. Smaller values of $\mu_H$ for $x$ = 0.5, 0.6 among all samples could be caused by the boundary scattering between the cubic and hexagonal phases.

Figure 6(b) represents the temperature dependence of $S$ of AgBiSe$_{2-2x}$S$_x$Te$_x$ with $x$ = 0–0.8. The negative sign of the $S$ indicates the n-type conduction for all the samples, and it is consistent with the Hall coefficient results. The temperature dependence of $S$ for pristine AgBiSe$_2$ exhibited the same trend as reported in Ref. 7. The value of $S$ was −138 μVK$^{-1}$ at room temperature, and it slightly increased until the $T_1$ (~500 K). In the rhombohedral phase, the $S$ value suddenly increased up to −300 μVK$^{-1}$ at around $T_2$ (~600 K), and it gradually decreased to −258 μVK$^{-1}$ at 748 K. The other hexagonal phase samples also showed almost similar behavior. In contrast, the samples above $x$ = 0.5 exhibited much higher $S$ value at room temperature, for instance, $S$ for $x$ = 0.6 sample was −256 μVK$^{-1}$ at room temperature. The $S$ of $x$ = 0.5 and 0.6 remained almost flat until 600 K, and it continuously decreased to −186 μVK$^{-1}$ at 748 K for both samples. Although $x$ = 0.7 and $x$ = 0.8 are also cubic at room temperature, the $S$ value of them exhibited a different behavior from the $x$ = 0.5 and 0.6 samples. It showed large reduction from around 500 K due to the creation of Bi$_2$Ch$_3$ impurity phase, which accords with the $\rho$ behavior. In general, the $\rho$ and the absolute value of $S$ are inversely proportional to $n_H$. This is consistent with the behavior of $S$ at room temperature in the hexagonal phase, whereas the absolute value of $S$ for cubic phase tended to decrease even though the $\rho$ increased. This should be affected by not only $n_H$ but also a decrease of $\mu_H$, which agrees with the smaller values of $\mu_H$ for the $x$ = 0.5, 0.6 samples among all of them (see Supplement Fig. S9), possibly due to the boundary scattering between the cubic and hexagonal phases.

Assuming a single parabolic band (SPB) model, the $S$ in degenerated semiconductors can be illustrated by the Pisarenko plot [33]. The $n_H$ dependence of $S$ is plotted in Figure 6(c). Each curve was generated at 300 K with an effective mass of $m^*/m_e$, where $m^*$ and $m_e$ are density of state effective mass and effective mass, equals to 1.2, 1.0, 0.8, 0.6, and 0.4 for obtained samples. The estimated $m^*/m_e$ for $x$ = 0.0–0.8 samples are 0.59, 0.77, 0.78, 0.58, 0.53, 1.25, 0.79, 0.56, and 0.43, respectively. The samples with $x = 0.0–0.4$ roughly followed the curves of the SPB model, indicating that the systematic changes of $\rho$ and $S$ with $x = 0.0–0.4$ could be mainly caused by the change of $n_H$. In contrast, the samples above $x = 0.5$ deviated from the SPB model trend, showing that the changes of $\rho$ and $S$ could be influenced by changes in band curvature or multiple bands [33,34]. Taking account of the different crystal structure for hexagonal and cubic structures, the band structure also differs from the both of them and it resulted in the different trends for $x = 0.0–0.4$ and $0.5–0.8$ samples.

Figure 6(d) presents the temperature dependence of power factors (PF) of AgBiSe$_{2-2x}$S$_x$Te$_x$ with $x = 0.0–0.8$. PF increased up to $T_1$ and decreased down to $T_2$ temperature for the $x = 0.0–0.4$ samples. In the cubic phase of all the samples, PF exhibited an increase trend as a function of temperature. While the maximum PF values were obtained at around $T_1$ region for $x = 0.0–0.2$ samples, the other samples showed highest values at highest temperature. The highest PF of 5.5 µWcm$^{-1}$K$^{-2}$ ($T = 485$ K) and 4.8 µWcm$^{-1}$K$^{-2}$ ($T = 747$ K) for the hexagonal and cubic phase were achieved for the $x = 0.0$ and $x = 0.3$ samples.

Figure 7(a) shows the temperature dependence of $\kappa_{tot}$ of AgBiSe$_{2-2x}$S$_x$Te$_x$ with $x = 0.0–0.8$. The pristine AgBiSe$_2$ exhibited relatively low $\kappa_{tot}$ value of 0.65 Wm$^{-1}$K$^{-1}$ at room temperature, and it slightly decreased until the $T_2$ (~598 K). Subsequently, a sudden decrease of $\kappa_{tot}$ down to 0.42 Wm$^{-1}$K$^{-1}$, possibly due to the structural transition from hexagonal/rhombohedral to cubic phase with an order–disorder transition of cation sites, was observed and then it gradually decreased with temperature. In the case of $x = 0.5$ sample, the behavior with order–disorder transition was not observed since the fraction of ASD was larger than those of the other hexagonal phase samples. Since any phase transition does not occur in $x = 0.6–0.8$ samples, the behavior was different from that for the hexagonal phase. For $x = 0.6$, the $\kappa_{tot}$ was 0.52 Wm$^{-1}$K$^{-1}$ at room temperature, and it slightly decreased to 0.42 Wm$^{-1}$K$^{-1}$ and remained almost flat against the temperature. For the $x = 0.7$ and 0.8 samples, a large increase of $\kappa_{tot}$ was observed from around 500 K possibly due to the emergence of the impurity phase of Bi$_2$Ch$_3$ phase, which has higher thermal conductivity such as typical thermoelectric material of Bi$_2$Te$_3$.

Fig. 7(b) represents the temperature dependence of $\kappa_{lat}$, subtracting the $\kappa_{ele}$ from $\kappa_{tot}$. The $\kappa_{ele}$ has been determined by using the Wiedemann−Franz law, $\kappa_{ele} = LT\rho^{-1}$, where $L$ is the Lorenz number. The $L$ is estimated using an equation $L = 1.5 + \exp(-|S|/116)$ [34]. The ultra-low $\kappa_{lat}$ was observed in all the samples; moreover, the $x = 0$–$0.7$ samples exhibited lower $\kappa_{lat}$ of 0.50 Wm$^{-1}$K$^{-1}$ above 620 K. The lowest $\kappa_{lat}$ of 0.24 Wm$^{-1}$K$^{-1}$ was obtained in $x = 0.4$ and 0.7 at around 750 K. From $T_1$ to $T_2$, humps in temperature dependence of $\kappa_{lat}$ were observed in the $x = 0.0$–$0.5$ samples. Although this kind of hump is rare, similar λ-shaped transition was also found in a few bulk crystals such as AgBiSe$_2$ order-disorder transition [36–38]. According to the previous reports, the ultra-low $\kappa_{lat}$ of AgBiSe$_2$ in the cubic phase is due to (a) phonon softening owing to a high degree of anharmonicity of Bi-Se bonds caused by the stereochemical active lone pair electrons (6s$^2$) of Bi, and (b) effective phonon scattering by the disordered Ag/Bi lattice [7,8,10]. In addition to these, effective phonon scattering should be increased by the disordered chalcogen framework with S, Se, and Te atoms, and it should result in the reduction tendency of $\kappa_{lat}$ around 360 K with increasing $x$. To clarify this reduction tendency of $\kappa_{lat}$, we discuss the decrease of $\kappa_{lat}$ with $x$ amount using the point defect scattering (PDS) model [8,9,39–47]. Figure 7(c) shows the calculated result of $\kappa_{lat}$ using the PDS model. This model can assess the $\kappa_{lat}$ of the contribution of mass and strain contrast. First, the $\kappa_{lat}$ of AgBiSe$_2$ was used as an initial value ($\kappa_{lat,pure}$). Average sound velocity of 1579.6 ms$^{-1}$, which obtained from longitudinal sound velocity of 2580.6 ms$^{-1}$ and transversal sound velocity of 1416.9 ms$^{-1}$, was in good agreement with the previous reports [8]. The elastic parameter $\varepsilon$ should be regarded as a phenomenological adjustable parameter, and the estimated $\varepsilon$ of 32.5 gives good fitting for the trend of measured room-temperature $\kappa_{lat}$. Following the model estimated with the contribution of mass and strain contrast is indicating the lattice disorder and/or strain by S and Te substitution fostered the phonon scattering and contributed to the decrease of $\kappa_{lat}$. Slightly lower $\kappa_{lat}$ from the model line could be explained by following 3 points, phonon scattering by increase of (1) an ASD and (2) Grüneisen parameter (see Supplemental Fig. S10), and (3) emergence of local chemical ordering (LCO) which will be described later. Although the qualitative estimation of contribution for $\kappa_{lat}$ from each point is difficult, above 3 points could contribute at least a small degree to the $\kappa_{lat}$ reduction.

In order to gain an insight of microstructures into the electrical and thermal transport characteristics, high-resolution TEM (HRTEM) observation was performed for the $x$ = 0, 0.2, 0.3, 0.4, 0.5, and 0.7 samples as shown in Figs. S11(a–f). For pristine AgBiSe$_2$, the atoms were clearly aligned (see Supplemental Fig. S11a) and the compositional segregation was not observed. On the other hand, although the atoms were also clearly aligned for the $x$ = 0.5 sample, the bright area and the dark area were observed (see Fig. 7d). According to the result of EDS analysis, localized aggregation of Ag and Te atoms, so called local chemical ordering (LCO) which is previously found in high-entropy alloys (HEAs) [48–50], was observed. The LCO was also observed in the $x$ = 0.7 sample (see Fig. S11f), whereas it was not observed in the $x$ = 0.0–0.4 samples. This finding of LCO suggests that an agglomeration of atoms might occur without occupying the cation and anion sites randomly in high-entropy materials of AgBiCh$_2$ similar to the HEAs case. The increase of ASD in hexagonal structure of $x$ = 0.0–0.4 would be also a sign of LCO occurrence.

Finally, temperature dependence of $ZT$ of AgBiSe$_{2-2x}$S$_x$Te$_x$ with $x$ = 0–0.8 are shown in Fig. 8 (a). Note that $ZT$ values plotted here are only for the cubic phases. A peak $ZT$ values of 0.62 at 748 K was achieved for pristine AgBiSe$_2$ sample. The $ZT$ values of all other samples except for $x$ = 0.8 surpassed that of pristine AgBiSe$_2$ above 660 K and the previously reported Sb-substituted cubic structural samples [14]. Among those samples, a peak $ZT$ value of 0.89 was achieved for $x$ = 0.3 at around 750 K. In the case of halogen doped AgBiSe$_2$, the peak value of AgBiSe$_{1.98}$Cl$_{0.02}$ was almost the same as that of $x$ = 0.3 sample; however, the absolute value of $S$ for $x$ = 0.3 sample was larger than that of AgBiSe$_{1.98}$Cl$_{0.02}$ [7], indicating that $ZT$ values of AgBiSe$_{2-2x}$S$_x$Te$_x$ samples have the possibility to exceed a unity by carrier tuning. Furthermore, the $ZT$ value of Nb-doped AgBiSe$_2$ (Ag$_{0.96}$Nb$_{0.04}$BiSe$_2$) reached 1 at 773 K [6]; nevertheless, $\kappa_{tot}$ was almost twice higher than that of AgBiSe$_{2-2x}$S$_x$Te$_x$ samples. In addition, the average $ZT$ values at 360–750 K for $x$ = 0.6 and 0.7 exhibited 0.38 and 0.40, which are comparable to the highest value of the average $ZT$ value reported in Br doped (AgBiSe$_2$)$_{0.7}$(PbSe)$_{0.3}$ [53]. Therefore, further investigation of carrier doped AgBiSe$_{2-2x}$S$_x$Te$_x$ could make us to achieve the higher $ZT$ > 1 and possibly the highest $ZT_{ave}$ among the AgBiSe$_2$-based materials.

**Conclusion**

In summary, we successfully stabilized the cubic AgBiSe$_{2-2x}$S$_x$Te$_x$ phase above $x = 0.6$ from room temperature to 800 K by utilizing the high-entropy concept through a simultaneous substitution of the Se site with S and Te. Ultra-low $\kappa_{lat}$ = 0.24 Wm$^{-1}$K$^{-1}$ was obtained with $x$ = 0.4 and 0.7 at 750 K. The decrease in $\kappa_{lat}$ was inversely proportional to the $x$ amount, indicating that effective point defect scattering was enhanced by ASD and the solid solution of the chalcogen atoms. Finally, a high peak $ZT$ value of 0.89 ($T$ = 750 K) was achieved. In addition, high average $ZT$ values of 0.38 and 0.40 were achieved between 360–750 K with $x$ = 0.6 and 0.7 without carrier tuning.

The present work demonstrates the stabilization of the cubic structure using the HE concept, together with achieving a high thermoelectric performance owing to the synergy between an ultra-low $\kappa_{lat}$ by effective point defect scattering and better transport properties by suppression of the bipolar effect.

**Credit author statements**

Asato Seshita: Formal analysis, Investigation, Resources, Data Curation, Writing -Original Draft, Writing - Review & Editing, Visualization.

Aichi Yamashita: Conceptualization, Methodology, Validation, Formal analysis, Investigation, Resources, Data Curation, Writing -Original Draft, Writing - Review & Editing, Visualization, Supervision, Project administration, Funding acquisition.

Takeshi Fujita: Investigation, Formal analysis, Resources, Data Curation, Writing - Review & Editing.

Takayoshi Katase: Investigation, Formal analysis, Resources, Data Curation, Writing - Review & Editing.

Akira Miura: Investigation, Writing - Review & Editing, Resources.

Yuki Nakahira: Investigation, Validation, Resources, Data Curation, Writing - Review & Editing.

Chikako Moriyoshi: Investigation, Data Curation.

Yoshihiro Kuroiwa: Investigation.

Yoshikazu Mizuguchi: Conceptualization, Writing - Review & Editing, Supervision, Project administration, Funding acquisition.

**Declaration of competing interest**

The authors declare that they have no known competing financial interests or personal relationships that could have appeared to influence the work reported in this paper.

**Data availability**

Data will be made available on request.

**Acknowledgements**

A. Y. was partly supported by a Grant-in-Aid for Scientific Research (KAKENHI) (No. 22K14480), and the Asahi Glass Foundation. Y. M. was partly supported by a Grant-in-Aid for Scientific Research (KAKENHI) (No. 21H00151), JST-ERATO (No. JPMJER2201), and the Tokyo Metropolitan Government Advanced Research (No. H31-1). We would like to thank Editage (www.editage.jp) for English language editing.

**Appendix A. Supplementary data**

Supplementary data to this article can be found online at

**References**

[1] G.J. Snyder, E.S. Toberer, Complex thermoelectric materials, Nat. Mater. **7** (2008) 105–114, https://doi.org/10.1038/nmat2090.

[2] T. Zhu, Y. Liu, C. Fu, J.P. Heremans, J.G. Snyder, X. Zhao, Compromise and Synergy in High-Efficiency Thermoelectric Materials, Adv. Mater. **29** (2017) 1605884, https://doi.org/10.1002/adma.201605884.

[3] L. Yang, Z.-G. Chen, M.S. Dargusch, J. Zou, High Performance Thermoelectric Materials: Progress and Their Applications, Adv. Energy Mater. **8** (2018) 1701797, https://doi.org/10.1002/aenm.201701797.

[4] D.S. Parker, A.F. May, D.J. Singh, Benefits of Carrier-Pocket Anisotropy to Thermoelectric Performance: The Case of p-Type $AgBiSe_2$, Phys. Rev. Appl. **3** (2015) 064003, https://doi.org/10.1103/PhysRevApplied.3.064003.

[5] C. Xiao, X. Qin, J. Zhang, R. An, J. Xu, K. Li, B. Cao, J. Yang, B. Ye, Y. Xie, High Thermoelectric and Reversible p-n-p Conduction Type Switching Integrated in Dimetal Chalcogenide, J. Am. Chem. Soc. **134** (2012) 18460–18466, https://doi.org/10.1021/ja308936b.

[6] L. Pan, D. Bérardan, N. Dragoe, High Thermoelectric Properties of n-Type $AgBiSe_2$, J. Am. Chem. Soc. **135** (2013) 4914–4917, https://doi.org/10.1021/ja312474n.

[7] S.N. Guin, V. Srihari, K. Biswas, Promising thermoelectric performance in n-type AgBiSe2: effect of aliovalent anion doping, J. Mater. Chem. A **3** (2014) 648–655,

https://doi.org/10.1039/C4TA04912H.

[8] F. Böcher, S.P. Culver, J. Peilstöcker, K.S. Weldert, W.G. Zeier, Vacancy and anti-site disorder scattering in AgBiSe$_2$ thermoelectrics, Dalton Trans. **46** (2017) 3906–3914, https://doi.org/10.1039/C7DT00381A.

[9] Y. Goto, A. Nishida, H. Nishiate, M. Murata, C.H. Lee, A. Miura, C. Moriyoshi, Y. Kuroiwa, Y. Mizuguchi, Effect of Te substitution on crystal structure and transport properties of AgBiSe$_2$ thermoelectric material, Dalton Trans. **47** (2018) 2575–2580, https://doi.org/10.1039/C7DT04821A.

[10] T. Bernges, J. Peilstöcker, M. Dutta, S. Ohno, S.P. Culver, K. Biswas, W.G. Zeier, Local Structure and Influence of Sb Substitution on the Structure–Transport Properties in AgBiSe$_2$, Inorg. Chem. 58 (2019) 9236–9245, https://doi.org/10.1021/acs.inorgchem.9b00874.

[11] S. Li, Z. Feng, Z. Tang, F. Zhang, F. Cao, X. Liu, D.J. Singh, J. Mao, Z. Ren, Q. Zhang, Defect Engineering for Realizing p-Type AgBiSe$_2$ with a Promising Thermoelectric Performance, Chem. Mater. **32** (2020) 3528–3536, https://doi.org/10.1021/acs.chemmater.0c00481.

[12] Y. Hu, S. Yuan, H. Huo, J. Xing, K. Guo, X. Yang, J. Luo, G.-H. Rao, J.-T. Zhao, Stabilized cubic phase BiAgSe$_{2-x}$S$_x$ with excellent thermoelectric properties via phase boundary engineering, J. Mater. Chem. C **9** (2021) 6766–6772, https://doi.org/10.1039/D1TC00760B.

[13] S. Mitra, D. Berardan, Influence of the temperature and composition on the crystal structure of the AgBiSe$_2$-AgBiS$_2$ system, Cryst. Res. Technol. **52** (2017) 1700075, https://doi.org/10.1002/crat.201700075.

[14] K. Sudo, Y. Goto, R. Sogabe, K. Hoshi, A. Miura, C. Moriyoshi, Y. Kuroiwa, Y. Mizuguchi, Doping-Induced Polymorph and Carrier Polarity Changes in Thermoelectric Ag(Bi,Sb)Se$_2$ Solid Solution, Inorg. Chem. **58** (2019) 7628–7633, https://doi.org/10.1021/acs.inorgchem.9b01038.

[15] J.-W. Yeh, S.-K. Chen, S.-J. Lin, J.-Y. Gan, T.-S. Chin, T.-T. Shun, C.-H. Tsau, S.-Y. Chang, Nanostructured High-Entropy Alloys with Multiple Principal Elements: Novel Alloy Design Concepts and Outcomes, Adv. Eng. Mater. **6** (2004) 299–303,

https://doi.org/10.1002/adem.200300567.

[16] J.-W. Yeh, Alloy Design Strategies and Future Trends in High-Entropy Alloys, JOM **65** (2013) 1759–1771, https://doi.org/10.1007/s11837-013-0761-6.

[17] A. Yamashita, Y. Goto, A. Miura, C. Moriyoshi, Y. Kuroiwa, Y. Mizuguchi, n-Type thermoelectric metal chalcogenide (Ag,Pb,Bi)(S,Se,Te) designed by multi-site-type high-entropy alloying, Mater. Res. Lett. **9** (2021) 366–372, https://doi.org/10.1080/21663831.2021.1929533.

[18] R. Sogabe, Y. Goto, Y. Mizuguchi, Superconductivity in REO$_{0.5}$F$_{0.5}$BiS$_2$ with high-entropy-alloy-type blocking layers, Appl. Phys. Express **11** (2018) 053102, https://doi.org/10.7567/APEX.11.053102.

[19] R. Sogabe, Y. Goto, T. Abe, C. Moriyoshi, Y. Kuroiwa, A. Miura, K. Tadanaga, Y. Mizuguchi, Improvement of superconducting properties by high mixing entropy at blocking layers in BiS$_2$-based superconductor REO$_{0.5}$F$_{0.5}$BiS$_2$, Solid State Commun. **295** (2019) 43–49, https://doi.org/10.1016/j.ssc.2019.04.001.

[20] Y. Mizuguchi, Superconductivity in High-Entropy-Alloy Telluride AgInSnPbBiTe$_5$, J. Phys. Soc. Jpn. **88** (2019) 124708, https://doi.org/10.7566/JPSJ.88.124708.

[21] M.R. Kasem, K. Hoshi, R. Jha, M. Katsuno, A. Yamashita, Y. Goto, T.D. Matsuda, Y. Aoki, Y. Mizuguchi, Superconducting properties of high-entropy-alloy tellurides M-Te (M: Ag, In, Cd, Sn, Sb, Pb, Bi) with a NaCl-type structure, Appl. Phys. Express **13** (2020) 033001, https://doi.org/10.35848/1882-0786/ab7482.

[22] Y. Shukunami, A. Yamashita, Y. Goto, Y. Mizuguchi, Synthesis of RE123 high-$T_c$ superconductors with a high-entropy-alloy-type RE site, Phys. C Supercond. Its Appl. **572** (2020) 1353623, https://doi.org/10.1016/j.physc.2020.1353623.

[23] A. Yamashita, R. Jha, Y. Goto, T.D. Matsuda, Y. Aoki, Y. Mizuguchi, An efficient way of increasing the total entropy of mixing in high-entropy-alloy compounds: a case of NaCl-type (Ag,In,Pb,Bi)Te$_{1-x}$Se$_x$ ($x$ = 0.0, 0.25, 0.5) superconductors, Dalton Trans. **49** (2020) 9118–9122, https://doi.org/10.1039/D0DT01880E.

[24] Y. Mizuguchi, Md.R. Kasem, T.D. Matsuda, Superconductivity in CuAl$_2$-type Co$_{0.2}$Ni$_{0.1}$Cu$_{0.1}$Rh$_{0.3}$Ir$_{0.3}$Zr$_2$ with a high-entropy-alloy transition metal site, Mater. Res. Lett. **9** (2021) 141–147, https://doi.org/10.1080/21663831.2020.1860147.

[25] Md.R. Kasem, A. Yamashita, Y. Goto, T.D. Matsuda, Y. Mizuguchi, Synthesis of high-entropy-alloy-type superconductors (Fe,Co,Ni,Rh,Ir)Zr$_2$ with tunable transition temperature, J. Mater. Sci. **56** (2021) 9499–9505, https://doi.org/10.1007/s10853-021-05921-2.

[26] B. Jiang, Y. Yu, J. Cui, X. Liu, L. Xie, J. Liao, Q. Zhang, Y. Huang, S. Ning, B. Jia, B. Zhu, S. Bai, L. Chen, S.J. Pennycook, J. He, High-entropy-stabilized chalcogenides with high thermoelectric performance, Science **371** (2021) 830–834, https://doi.org/10.1126/science.abe1292.

[27] B. Jiang, W. Wang, S. Liu, Y. Wang, C. Wang, Y. Chen, L. Xie, M. Huang, J. He, High figure-of-merit and power generation in high-entropy GeTe-based thermoelectrics, Science **377** (2022) 208–213, https://doi.org/10.1126/science.abq5815.

[28] Z. Guo, Y.K. Zhu, M. Liu, X. Dong, B. Sun, F. Guo, Q. Zhang, J. Li, W. Gao, Y. Fu, W. Cai, J. Sui, Z. Liu, Cubic phase stabilization and thermoelectric performance optimization in AgBiSe$_2$–SnTe system, Mater. Today Phys. **38**, (2023) 101238, https://doi.org/10.1016/j.mtphys.2023.101238.

[29] S. Kawaguchi, M. Takemoto, K. Osaka, E. Nishibori, C. Moriyoshi, Y. Kubota, Y. Kuroiwa, K. Sugimoto, High-throughput powder diffraction measurement system consisting of multiple MYTHEN detectors at beamline BL02B2 of SPring-8, Rev. Sci. Instrum. **88** (2017) 085111. https://doi.org/10.1063/1.4999454.

[30] F. Izumi, K. Momma, Three-Dimensional Visualization in Powder Diffraction, Solid State Phenom. **130** (2007) 15–20, https://doi.org/10.4028/www.scientific.net/SSP.130.15.

[31] K. Momma, F. Izumi, VESTA 3 for three-dimensional visualization of crystal, volumetric and morphology data, J. Appl. Crystallogr. **44** (2011) 1272–1276, https://doi.org/10.1107/S0021889811038970.

[32] Y. Zhang, T.T. Zuo, Z. Tang, M.C. Gao, K.A. Dahmen, P.K. Liaw, Z.P. Lu, Microstructures and properties of high-entropy alloys, Prog. Mater. Sci. **61** (2014) 1–93, https://doi.org/10.1016/j.pmatsci.2013.10.001.

[33] A.F. May, G.J. Snyder, Introduction to Modeling Thermoelectric Transport at High Temperatures, (2012) 1–18.

[34] Y. Luo, S. Hao, S. Cai, T.J. Slade, Z.Z. Luo, V.P. Dravid, C. Wolverton, Q. Yan, M.G. Kanatzidis, High Thermoelectric Performance in the New Cubic Semiconductor $AgSnSbSe_3$ by High-Entropy Engineering, J. Am. Chem. Soc. **142** (2020) 15187–15198, https://doi.org/10.1021/jacs.0c07803.

[35] H.-S. Kim, Z.M. Gibbs, Y. Tang, H. Wang, G.J. Snyder, Characterization of Lorenz number with Seebeck coefficient measurement, APL Mater. **3** (2015) 041506, https://doi.org/10.1063/1.4908244.

[36] F. Gascoin, A. Maignan, Order–Disorder Transition in $AgCrSe_2$: a New Route to Efficient Thermoelectrics, Chem. Mater. **23** (2011) 2510–2513, https://doi.org/10.1021/cm200581k.

[37] C. Xiao, J. Xu, K. Li, J. Feng, J. Yang, Y. Xie, Superionic Phase Transition in Silver Chalcogenide Nanocrystals Realizing Optimized Thermoelectric Performance, J. Am. Chem. Soc. **134** (2012) 4287–4293, https://doi.org/10.1021/ja2104476.

[38] S.N. Guin, K. Biswas, Cation Disorder and Bond Anharmonicity Optimize the Thermoelectric Properties in Kinetically Stabilized Rocksalt $AgBiS_2$ Nanocrystals, Chem. Mater. **25** (2013) 3225–3231, https://doi.org/10.1021/cm401630d.

[39] B. Abeles, Lattice Thermal Conductivity of Disordered Semiconductor Alloys at High Temperatures, Phys. Rev. **131** (1963) 1906–1911, https://doi.org/10.1103/PhysRev.131.1906.

[40] J. Yang, G.P. Meisner, L. Chen, Strain field fluctuation effects on lattice thermal conductivity of ZrNiSn-based thermoelectric compounds, Appl. Phys. Lett. **85** (2004) 1140–1142, https://doi.org/10.1063/1.1783022.

[41] P.G. Klemens, The Scattering of Low-Frequency Lattice Waves by Static Imperfections, Proc. Phys. Soc. Sect. A **68** (1955) 1113, https://doi.org/10.1088/0370-1298/68/12/303.

[42] J. Callaway, Model for Lattice Thermal Conductivity at Low Temperatures, Phys. Rev. **113** (1959) 1046–1051, https://doi.org/10.1103/PhysRev.113.1046.

[43] J. Callaway, H.C. von Baeyer, Effect of Point Imperfections on Lattice Thermal Conductivity, Phys. Rev. **120** (1960) 1149–1154, https://doi.org/10.1103/PhysRev.120.1149.

[44] O.L. Anderson, A simplified method for calculating the debye temperature from elastic constants, J. Phys. Chem. Solids **24** (1963) 909–917, https://doi.org/10.1016/0022-3697(63)90067-2.

[45] D.S. Sanditov, V.N. Belomestnykh, Relation between the parameters of the elasticity theory and averaged bulk modulus of solids, Tech. Phys. **56** (2011) 1619–1623, https://doi.org/10.1134/S106378421111020X.

[46] A. Zevalkink, E.S. Toberer, W.G. Zeier, E. Flage-Larsen, G.J. Snyder, $Ca_3AlSb_3$: an inexpensive, non-toxic thermoelectric material for waste heat recovery, Energy Environ. Sci. **4** (2011) 510–518, https://doi.org/10.1039/C0EE00517G.

[47] H. Wang, J. Wang, X. Cao, G.J. Snyder, Thermoelectric alloys between PbSe and PbS with effective thermal conductivity reduction and high figure of merit, J. Mater. Chem. A **2** (2014) 3169, https://doi.org/10.1039/c3ta14929c.

[48] H.-H. Wu, L.-S. Dong, S.-Z. Wang, G.-L. Wu, J.-H. Gao, X.-S. Yang, X.-Y. Zhou, X.-P. Mao, Local chemical ordering coordinated thermal stability of nanograined high-entropy alloys, Rare Met. **42** (2023) 1645–1655, https://doi.org/10.1007/s12598-022-02194-9.

[49] K. Yan, Y. Xu, J. Niu, Y. Wu, Y. Li, B. Gault, S. Zhao, X. Wang, Y. Li, J. Wang, K.P. Skokov, O. Gutfleisch, H. Wu, D. Jiang, Y. He, C. Jiang, Unraveling the origin of local chemical ordering in Fe-based solid-solutions, Acta Mater. **264** (2024) 119583, https://doi.org/10.1016/j.actamat.2023.119583.

[50] Z. Huang, T. Li, B. Li, Q. Dong, J. Smith, S. Li, L. Xu, G. Wang, M. Chi, L. Hu, Tailoring Local Chemical Ordering via Elemental Tuning in High-Entropy Alloys, J. Am. Chem. Soc. **146** (2024) 2167–2173, https://doi.org/10.1021/jacs.3c12048.

[51] T. Zhao, H. Zhu, B. Zhang, S. Zheng, N. Li, G. Wang, G. Wang, X. Lu, X. Zhou, High

thermoelectric performance of tellurium-free n-type AgBi$_{1-x}$Sb$_x$Se$_2$ with stable cubic structure enabled by entropy engineering, Acta Mater. **220** (2021) 117291, https://doi.org/10.1016/j.actamat.2021.117291.

[52] L. Zhang, W. Shen, Z. Zhang, C. Fang, Q. Wang, B. Wan, L. Chen, Y. Zhang, X. Jia, Stabilizing n-type cubic AgBiSe$_2$ thermoelectric materials through alloying with PbS, J. Materiomics **10** (2024) 70–77, https://doi.org/10.1016/j.jmat.2023.04.008.

[53] H. Zhu, T. Zhao, B. Zhang, Z. An, S. Mao, G. Wang, X. Han, X. Lu, J. Zhang, X. Zhou, Entropy Engineered Cubic n-Type AgBiSe$_2$ Alloy with High Thermoelectric Performance in Fully Extended Operating Temperature Range, Adv. Energy Mater. **11** (2021) 2003304, https://doi.org/10.1002/aenm.202003304.

**Figures and captions**

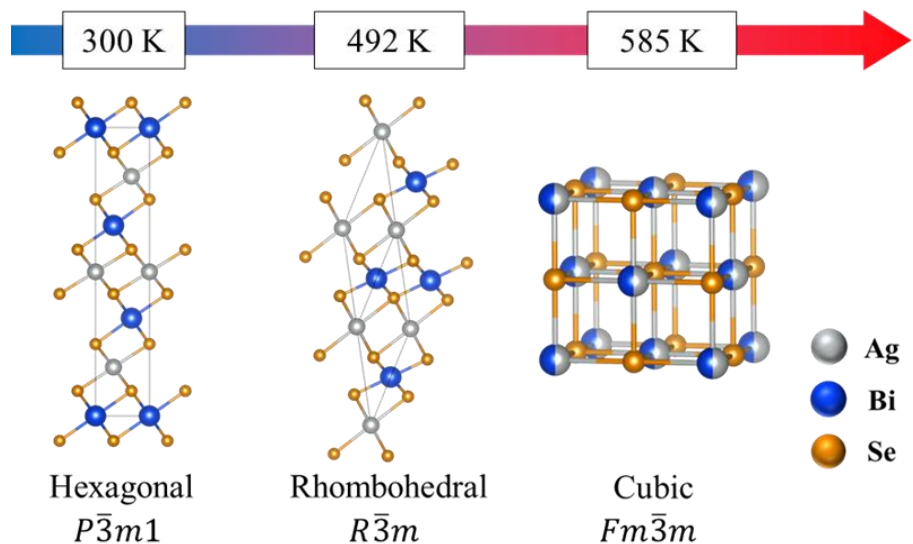

**Fig.1** Schematic images of crystal structure of AgBiSe$_2$ and its crystal structural transition against temperature.

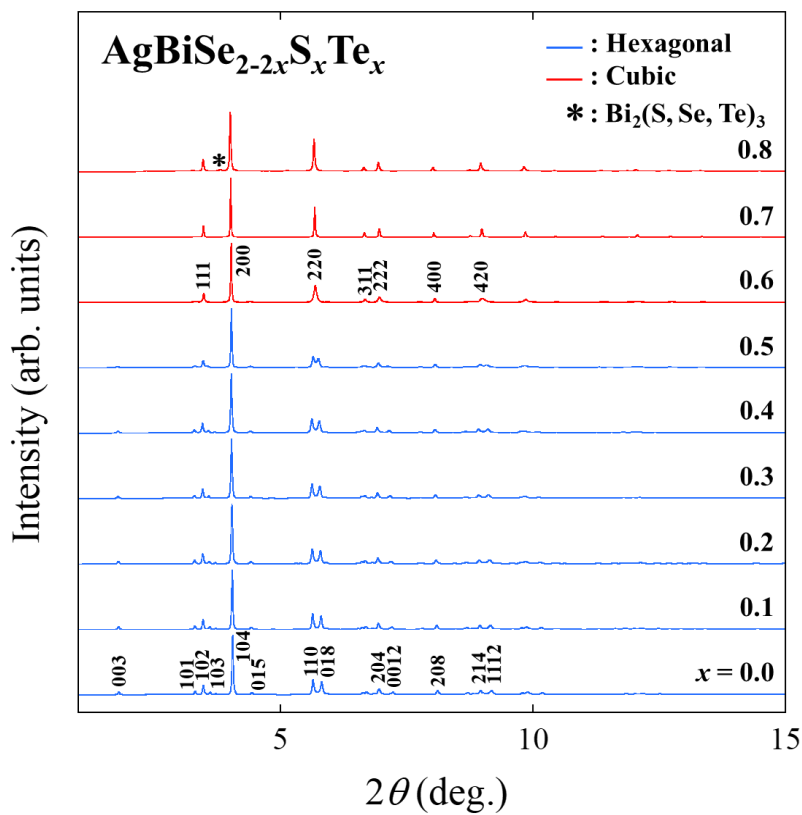

**Fig. 2** Synchrotron powder X-ray diffraction patterns of AgBiSe$_{2-2x}$S$_x$Te$_x$ with $x$ = 0.0–0.8 at room

temperature. The asterisk represents the diffraction peak due to the $Bi_2(S,Se,Te)_3$ secondary phase. The wavelength of the beam was determined to be $\lambda = 0.206684$ Å.

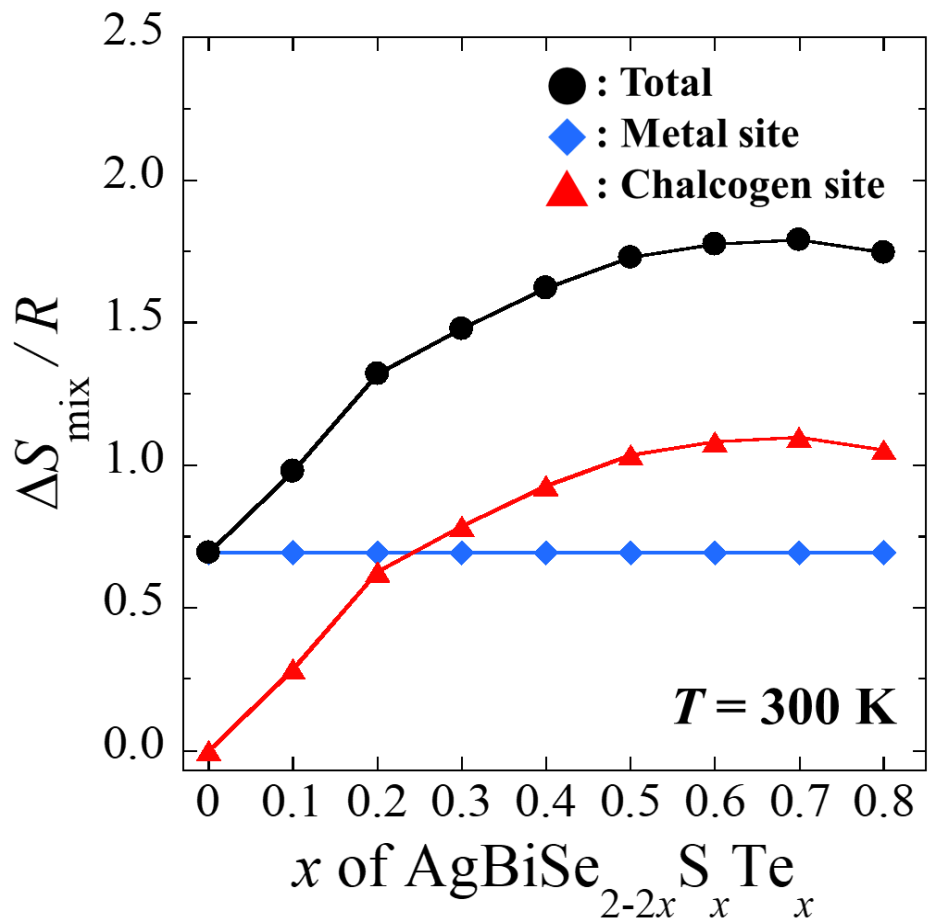

**Fig. 3** Estimated $\Delta S_{mix}/R$ as a function of $x$ amount.

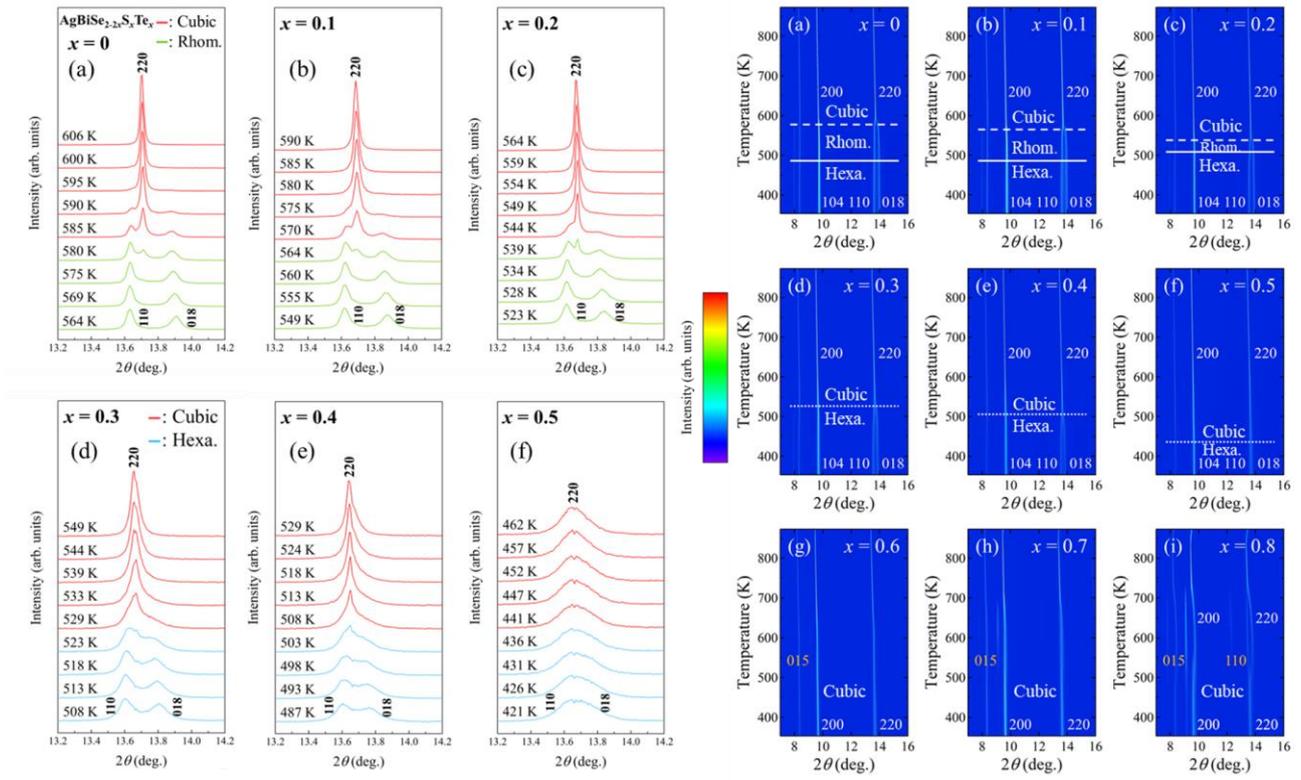

**Fig. 4** (a–f) Temperature dependence of synchrotron powder X-ray diffraction (SPXRD) patterns around (110) and (018) reflections of hexagonal structure and (220) reflection of cubic structure with $x$ = 0–0.5 samples. (g–l) 2D color mapping for temperature dependence of SPXRD patterns of AgBiSe$_{2-2x}$S$_x$Te$_x$ with $x$ = 0–0.8 samples.

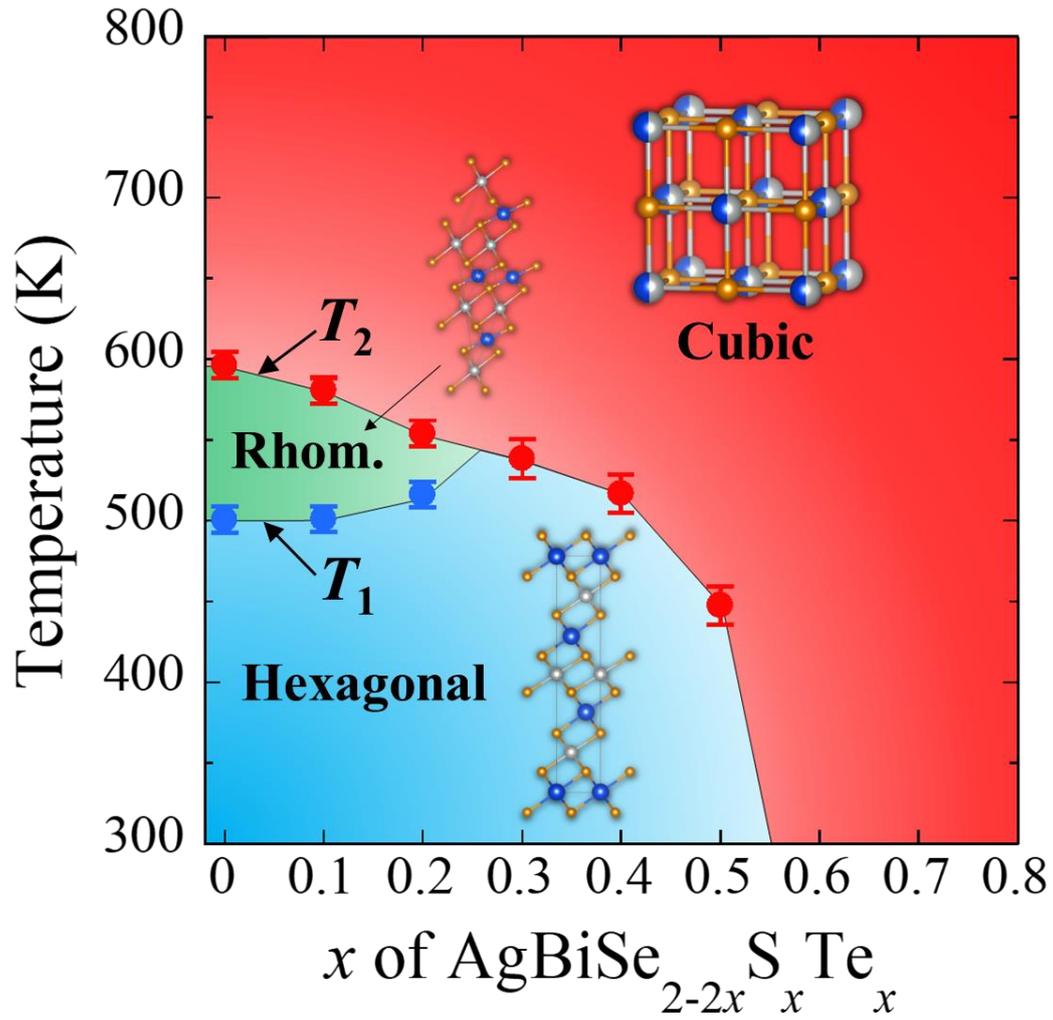

**Fig. 5** Temperature against *x* amount (*T-x*) phase diagram of AgBiSe$_{2-2x}$S$_x$Te$_x$ with $x = 0$–$0.8$. Note that $T_1$ and $T_2$ denote the structural transition temperature from hexagonal to rhombohedral structure and rhombohedral to cubic structure, respectively. Blue, green, and red regions indicate the dominant phase of hexagonal, rhombohedral, and cubic, respectively.

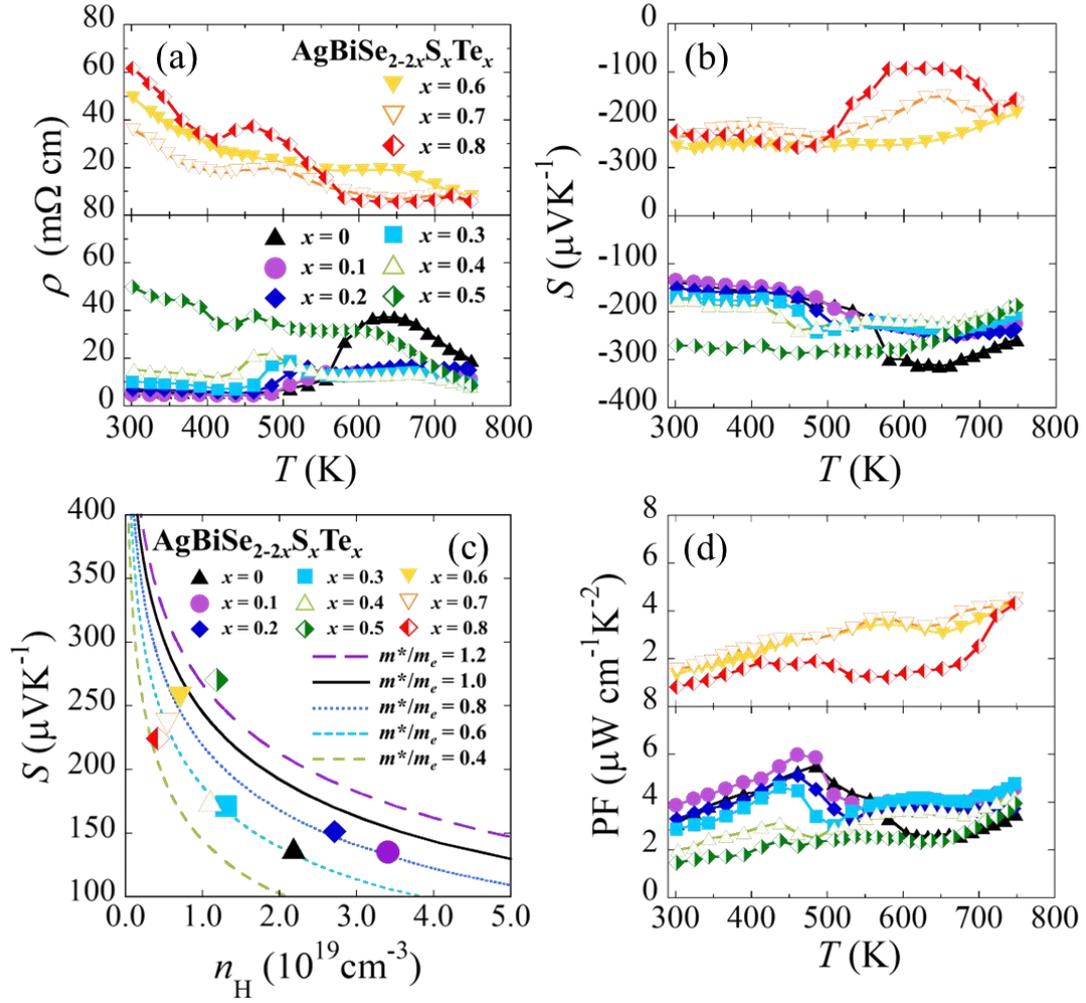

**Fig. 6** Temperature dependence of (a) electrical resistivity, (b) Seebeck coefficient, (c) Seebeck coefficient against carrier concentration known as Pisarenko plot, and (d) power factor of AgBiSe$_{2-2x}$S$_x$Te$_x$ with $x$ = 0–0.8.

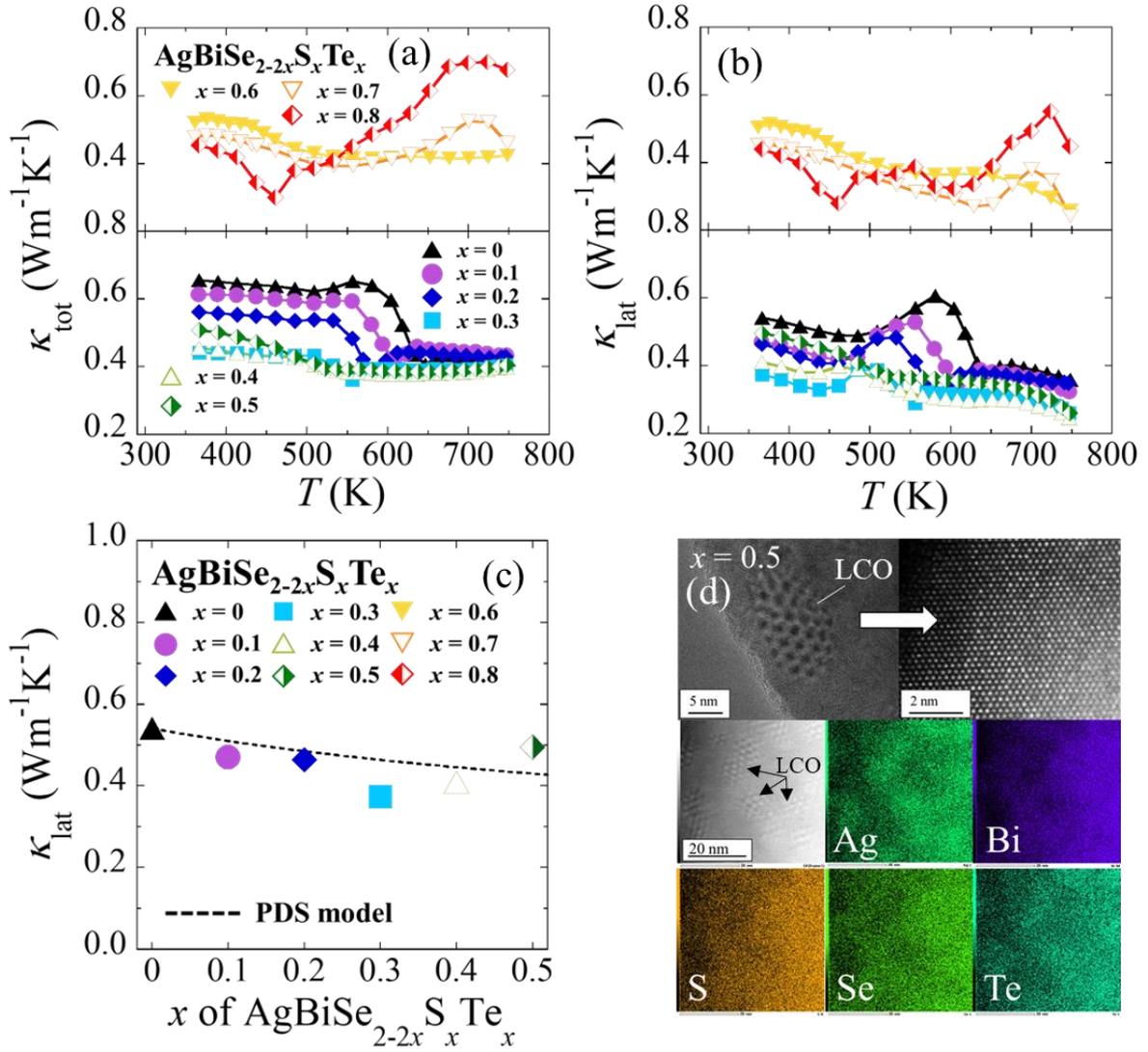

**Fig. 7** (a) Total thermal conductivity and (b) lattice thermal conductivity of AgBiSe$_{2-2x}$S$_x$Te$_x$ with $x$ = 0–0.8 against temperature. (c) Temperature dependence of lattice thermal conductivity at room temperature. (d) High resolution transmission electron microscopy (TEM) and scanning TEM (STEM) images and energy dispersive X-ray spectroscopy (EDS) mapping of AgBiSeS$_{0.5}$Te$_{0.5}$. Note that, arrows indicate local chemical ordering (LCO).

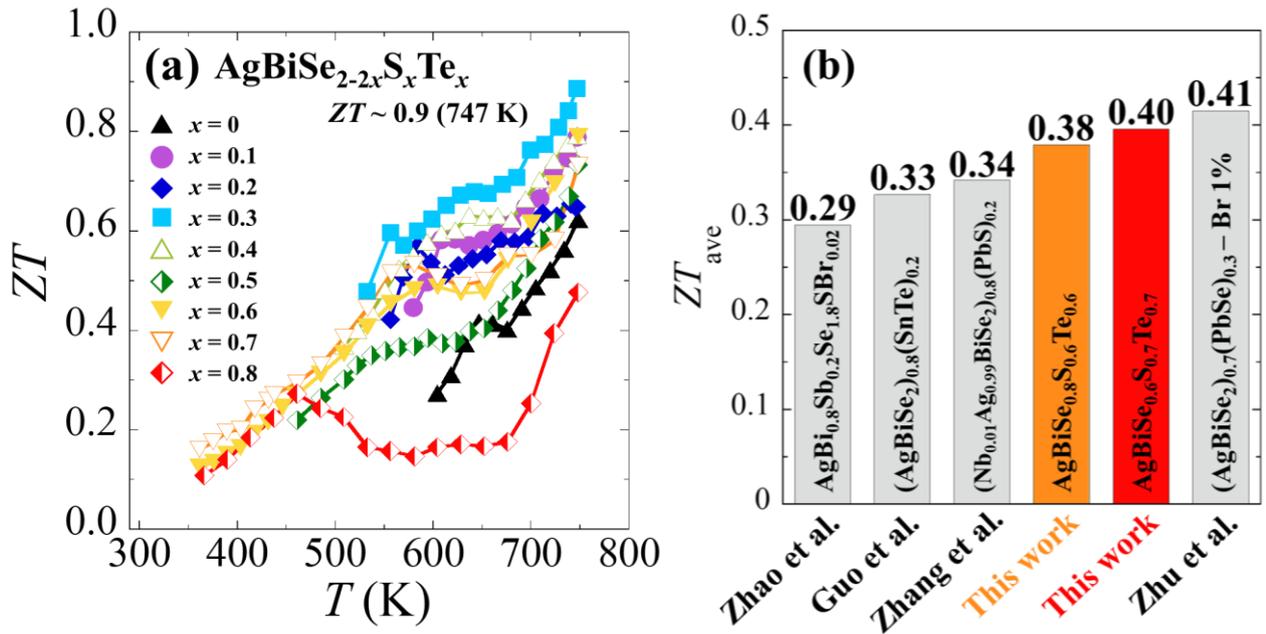

**Fig. 8** (a) Temperature dependence of dimensionless figure-of-merit ($ZT$) of AgBiSe$_{2-2x}$S$_x$Te$_x$ with $x$ = 0–0.8. (b) $ZT_{ave}$ value between 360 and 750 K for cubic structure of AgBiSe$_2$-based materials [28, 51–53].

# Supplemental information

**Tables**

**Table S1** Room temperature carrier concentration ($n_H$) and carrier mobility ($\mu_H$) of AgBiSe$_{2-2x}$S$_x$Te$_x$ with $x$ = 0–0.8.

| nominal | sample | $\Delta S_{metal}$ | $\Delta S_{chalcogen}$ | $\Delta S_{total}$ |
|---|---|---|---|---|
| AgBiSe$_2$ | Ag$_{1.02}$Bi$_{1.03}$Se$_{1.96}$ | 0.693$R$ | 0$R$ | 0.693$R$ |
| AgBiSe$_{1.8}$S$_{0.1}$Te$_{0.1}$ | Ag$_{1.05}$Bi$_{1.03}$Se$_{1.80}$S$_{0.03}$Te$_{0.11}$ | 0.693$R$ | 0.285$R$ | 0.978$R$ |
| AgBiSe$_{1.6}$S$_{0.2}$Te$_{0.2}$ | Ag$_{1.09}$Bi$_{1.09}$Se$_{1.47}$S$_{0.13}$Te$_{0.23}$ | 0.693$R$ | 0.627$R$ | 1.320$R$ |
| AgBiSe$_{1.4}$S$_{0.3}$Te$_{0.3}$ | Ag$_{1.05}$Bi$_{1.03}$Se$_{1.38}$S$_{0.24}$Te$_{0.30}$ | 0.693$R$ | 0.786$R$ | 1.479$R$ |
| AgBiSe$_{1.2}$S$_{0.4}$Te$_{0.4}$ | Ag$_{1.04}$Bi$_{1.02}$Se$_{1.20}$S$_{0.34}$Te$_{0.40}$ | 0.693$R$ | 0.928$R$ | 1.621$R$ |
| AgBiSe$_{1.0}$S$_{0.5}$Te$_{0.5}$ | Ag$_{1.04}$Bi$_{1.03}$Se$_{0.98}$S$_{0.45}$Te$_{0.51}$ | 0.693$R$ | 1.035$R$ | 1.728$R$ |
| AgBiSe$_{0.8}$S$_{0.6}$Te$_{0.6}$ | Ag$_{1.05}$Bi$_{1.01}$Se$_{0.81}$S$_{0.54}$Te$_{0.59}$ | 0.693$R$ | 1.082$R$ | 1.774$R$ |
| AgBiSe$_{0.6}$S$_{0.7}$Te$_{0.7}$ | Ag$_{1.03}$Bi$_{1.03}$Se$_{0.61}$S$_{0.64}$Te$_{0.70}$ | 0.693$R$ | 1.097$R$ | 1.790$R$ |
| AgBiSe$_{0.4}$S$_{0.8}$Te$_{0.8}$ | Ag$_{1.05}$Bi$_{1.02}$Se$_{0.38}$S$_{0.75}$Te$_{0.80}$ | 0.693$R$ | 1.053$R$ | 1.746$R$ |

**Table S2** Chemical composition, carrier type, structure type, $\Delta S_{mix}/R$ (Total), representative thermoelectric properties of maximum dimensionless figure-of-merit $ZT$ and PF and lowest kL at presented temperature, and $n_H$ and mobility at room temperature of the medium-entropy- and high-entropy-type thermoelectric materials.

| HE-type thermoelectric materials | carrier type | structure | $\Delta S_{mix}/R$ | T (K) | ZT | PF ($\mu W cm^{-1}K^{-2}$) | $\kappa_{lat}$ ($Wm^{-1}K^{-1}$) | $n_H$ @RT ($10^{18}cm^{-3}$) | $\mu_H$ @RT ($cm^2V^{-1}s^{-1}$) | reference |
|---|---|---|---|---|---|---|---|---|---|---|
| $AgBiSe_2$ | n | $Fm\text{-}3m$ | 0.69 | 748.3 | 0.62 | 3.50 | 0.36 | 21.82 (hexagonal) | 46.6 (hexagonal) | this work |
| $AgBiSe_{1.8}Sn_{0.1}Te_{0.1}$ | n | $Fm\text{-}3m$ | 0.98 | 747.4 | 0.79 | 4.56 | 0.32 | 34.11 (hexagonal) | 38.9 (hexagonal) | |
| $AgBiSe_{1.6}Sn_{0.2}Te_{0.2}$ | n | $Fm\text{-}3m$ | 1.32 | 745.9 | 0.65 | 3.73 | 0.35 | 27.14 (hexagonal) | 33.2 (hexagonal) | |
| $AgBiSe_{1.4}Sn_{0.3}Te_{0.3}$ | n | $Fm\text{-}3m$ | 1.48 | 747.3 | 0.89 | 4.78 | 0.26 | 13.06 (hexagonal) | 46.7 (hexagonal) | |
| $AgBiSe_{1.2}Sn_{0.4}Te_{0.4}$ | n | $Fm\text{-}3m$ | 1.62 | 747.2 | 0.79 | 4.22 | 0.24 | 11.03 (hexagonal) | 37.1 (hexagonal) | |
| $AgBiSeSn_{0.5}Te_{0.5}$ | n | $Fm\text{-}3m$ | 1.73 | 747.8 | 0.73 | 3.96 | 0.26 | 11.82 (hexagonal) | 10.6 (hexagonal) | |
| $AgBiSe_{0.8}Sn_{0.6}Te_{0.6}$ | n | $Fm\text{-}3m$ | 1.77 | 748.3 | 0.79 | 4.48 | 0.26 | 7.09 | 17.9 | |
| $AgBiSe_{0.6}Sn_{0.7}Te_{0.7}$ | n | $Fm\text{-}3m$ | 1.79 | 748.3 | 0.73 | 4.50 | 0.24 | 5.47 | 31.8 | |
| $AgBiSe_{0.4}Sn_{0.8}Te_{0.8}$ | n | $Fm\text{-}3m$ | 1.75 | 748.1 | 0.48 | 4.31 | 0.45 | 4.21 | 24.0 | |
| $Ag_{0.25}Pb_{0.50}Bi_{0.25}Sn_{0.40}Se_{0.50}Te_{0.10}$ | n | $Fm\text{-}3m$ | 2.00 | 723 | 0.54 | 4.40 | 0.46 | 80.4 | 18.5 | [1] |
| $AgSnSbSe_3$ | p | $Fm\text{-}3m$ | 1.10 | 773 | 0.7 | 5.2 | 0.5 | 28.3 | 20.5 | |
| $AgSnSbSe_{2.5}Te_{0.5}$ | p | $Fm\text{-}3m$ | 1.55 | 723 | 0.9 | 6.6 | 0.4 | 35.7 | 18.1 | |
| $AgSnSbSe_2Te$ | p | $Fm\text{-}3m$ | 1.74 | 773 | 0.9 | 7.3 | 0.4 | 47.2 | 20.1 | |
| $AgSnSbSe_{1.7}Te_{1.3}$ | p | $Fm\text{-}3m$ | 1.78 | 723 | 1.1 | 9.5 | 0.3 | 59.4 | 23.8 | |
| $AgSnSbSe_{1.5}Te_{1.5}$ | p | $Fm\text{-}3m$ | 1.79 | 723 | 1.1 | 9.9 | 0.3 | 65.7 | 27.6 | [2] |
| $AgSnSbSe_{1.3}Te_{1.7}$ | p | $Fm\text{-}3m$ | 1.78 | 673 | 1.1 | 9.9 | 0.4 | 70.5 | 28.2 | |
| $AgSnSbSeTe_2$ | p | $Fm\text{-}3m$ | 1.74 | 673 | 1.1 | 10.7 | 0.3 | 83.6 | 29.6 | |
| $AgSnSbSe_{0.5}Te_{2.5}$ | p | $Fm\text{-}3m$ | 1.55 | 723 | 0.9 | 10.8 | 0.4 | 97.3 | 29.6 | |
| $AgSnSbSeTe_3$ | p | $Fm\text{-}3m$ | 1.10 | 773 | 0.8 | 10.4 | 0.5 | 119.5 | 30.0 | |
| $Pb_{0.99}Sb_{0.012}Se_{0.5}Te_{0.25}S_{0.25}$ | n | $Fm\text{-}3m$ | 1.10 | 900 | 1.4 | 12.1 | 0.4 | 52 | 192.5 | |
| $Pb_{0.94}Sb_{0.012}Sn_{0.05}Se_{0.5}Te_{0.25}S_{0.25}$ | n | $Fm\text{-}3m$ | 1.30 | 900 | 1.5 | 12.4 | 0.3 | 44 | 182.4 | |
| $Pb_{0.89}Sb_{0.012}Sn_{0.1}Se_{0.5}Te_{0.25}S_{0.25}$ | n | $Fm\text{-}3m$ | 1.43 | 900 | 1.8 | 12.6 | 0.3 | 38 | 193.0 | [3] |
| $Pb_{0.84}Sb_{0.012}Sn_{0.15}Se_{0.5}Te_{0.25}S_{0.25}$ | n | $Fm\text{-}3m$ | 1.52 | 900 | 1.4 | 11.2 | 0.4 | 34 | 221.1 | |
| $Pb_{0.79}Sb_{0.012}Sn_{0.2}Se_{0.5}Te_{0.25}S_{0.25}$ | n | $Fm\text{-}3m$ | 1.60 | 800 | 1.1 | 9.6 | 0.5 | 33 | 162.8 | |
| $Ge_{0.77}Ag_{0.11}Pb_{0.12}Te$ | p | $Fm\text{-}3m$ | 0.70 | 800 | 0.6 | 24.0 | 0.8 | - | - | |
| $Ge_{0.75}Ag_{0.13}Pb_{0.12}Te$ | p | $Fm\text{-}3m$ | 0.74 | 800 | 1.1 | 15.1 | 0.9 | - | - | |
| $Ge_{0.74}Ag_{0.11}Pb_{0.15}Te$ | p | $Fm\text{-}3m$ | 0.73 | 800 | 1.7 | 33.9 | 0.4 | - | - | |
| $Ge_{0.62}Ag_{0.11}Sb_{0.15}Pb_{0.12}Te$ | p | $Fm\text{-}3m$ | 1.06 | 800 | 2.5 | 29.8 | 0.2 | - | - | |
| $Ge_{0.61}Ag_{0.11}Sb_{0.15}Pb_{0.12}Bi_{0.01}Te$ | p | $Fm\text{-}3m$ | 1.11 | 800 | 2.7 | 28.4 | 0.3 | - | - | |
| $Ge_{0.61}Ag_{0.11}Sb_{0.15}Pb_{0.12}Bi_{0.01}Cd_{0.05}Te$ | p | $Fm\text{-}3m$ | 1.13 | 750 | 2.3 | 25.1 | 0.4 | - | - | [4] |
| $Ge_{0.56}Ag_{0.11}Sb_{0.15}Pb_{0.12}Cd_{0.05}Bi_{0.01}Te$ | p | $Fm\text{-}3m$ | 1.29 | 800 | 1.9 | 19.6 | 0.5 | - | - | |
| $Ge_{0.56}Ag_{0.11}Sb_{0.15}Pb_{0.12}Mn_{0.05}Bi_{0.01}Te$ | p | $Fm\text{-}3m$ | 1.29 | 800 | 2.7 | 25.8 | 0.3 | - | - | |
| $Ge_{0.56}Ag_{0.11}Sb_{0.15}Pb_{0.12}Sn_{0.05}Bi_{0.01}Te$ | p | $Fm\text{-}3m$ | 1.29 | 800 | 2.5 | 24.5 | 0.3 | - | - | |
| $Pb_{0.975}Na_{0.025}Se_{0.5}S_{0.25}Te_{0.25}$ | p | $Fm\text{-}3m$ | 1.16 | 900 | 1.3 | 13.2 | 0.4 | 28 | 40 | |
| $Pb_{0.974}Cd_{0.01}Na_{0.025}Se_{0.5}S_{0.25}Te_{0.25}$ | p | $Fm\text{-}3m$ | 1.21 | 900 | 1.7 | 15.2 | 0.4 | 27 | 37 | |
| $Pb_{0.973}Cd_{0.02}Na_{0.025}Se_{0.5}S_{0.25}Te_{0.25}$ | p | $Fm\text{-}3m$ | 1.26 | 900 | 1.7 | 16.1 | 0.4 | 26 | 42 | [5] |
| $Pb_{0.972}Cd_{0.03}Na_{0.025}Se_{0.5}S_{0.25}Te_{0.25}$ | p | $Fm\text{-}3m$ | 1.29 | 900 | 1.9 | 15.4 | 0.4 | 23 | 36 | |
| $Pb_{0.971}Cd_{0.04}Na_{0.025}Se_{0.5}S_{0.25}Te_{0.25}$ | p | $Fm\text{-}3m$ | 1.32 | 900 | 2.0 | 16.2 | 0.4 | 29 | 41 | |
| $Pb_{0.97}Cd_{0.05}Na_{0.025}Se_{0.5}S_{0.25}Te_{0.25}$ | p | $Fm\text{-}3m$ | 1.36 | 900 | 1.8 | 14.8 | 0.4 | 21 | 34 | |
| $AgMnSnSbTe_4$ | p | $Fm\text{-}3m$ | 1.39 | 773 | 0.7 | 11.1 | 0.7 | 150 | 8 | |
| $AgMnSn_{0.75}Pb_{0.25}SbTe_4$ | p | $Fm\text{-}3m$ | 1.53 | 773 | 0.8 | 11.1 | 0.7 | 150 | 13 | |
| $AgMnSn_{0.5}Pb_{0.5}SbTe_4$ | p | $Fm\text{-}3m$ | 1.56 | 773 | 1.1 | 13.6 | 0.6 | 120 | 28 | [6] |
| $AgMnSn_{0.25}Pb_{0.75}SbTe_4$ | p | $Fm\text{-}3m$ | 1.53 | 773 | 1.3 | 11.0 | 0.5 | 70 | 31 | |
| $AgMnPbSbTe_4$ | p | $Fm\text{-}3m$ | 1.39 | 700 | 1.1 | 7.2 | 0.4 | 60 | 32 | |
| $CuInTe_2$ | p | $I\text{-}42d$ | 0.69 | 820 | 0.5 | 13.5 | 1.8 | - | - | |
| $Cu_{0.5}Ag_{0.5}(ZnGe)_{0.1}(GaIn)_{0.4}Te_2$ | p | $I\text{-}42d$ | 1.54 | 820 | 1.02 | 6.3 | 0.4 | 32 | 17.8 | |
| $Cu_{0.5}Ag_{0.5}(ZnGe)_{0.2}(GaIn)_{0.3}Te_2$ | p | $I\text{-}42d$ | 1.63 | 820 | 0.7 | 5.0 | 0.4 | 60 | 16.0 | [7] |
| $Cu_{0.5}Ag_{0.5}(ZnGe)_{0.25}(GaIn)_{0.25}Te_2$ | p | $I\text{-}42d$ | 1.64 | 820 | 0.7 | 5.1 | 0.5 | 73 | 15.9 | |
| $Cu_{0.5}Ag_{0.4}(ZnGe)_{0.25}(GaIn)_{0.25}Te_2$ | p | $I\text{-}42d$ | 1.73 | 820 | 0.4 | 2.3 | 0.4 | 56 | 1.2 | |
| $Cu_2Se_2S$ | p | $Cc$ | 1.27 | 873 | 0.8 | 37 | 0.3 | - | - | |
| $Cu_2(In_{0.05}Sn_{0.95})Se_2S$ | p | $Cc$ | 1.34 | 873 | 1.1 | 46 | 0.3 | - | - | |
| $Cu_2(In_{0.06}Sn_{0.94})Se_2S$ | p | $Cc$ | 1.35 | 873 | 1.2 | 41 | 0.2 | 65 | 1.75 | [8] |
| $Cu_2(In_{0.25}Sn_{0.75})Se_2S$ | p | $Cc$ | 1.46 | 873 | 0.9 | 35 | 0.3 | - | - | |
| $Cu_{1.95}Ag_{0.05}(In_{0.06}Sn_{0.94})Se_2S$ | p | $Cc$ | 1.45 | 873 | 1.4 | 46 | 0.2 | - | - | |
| $Cu_{1.87}Ag_{0.13}(In_{0.06}Sn_{0.94})Se_2S$ | p | $Cc$ | 1.51 | 873 | 1.52 | 44 | 0.1 | - | - | |

**Figures and captions**

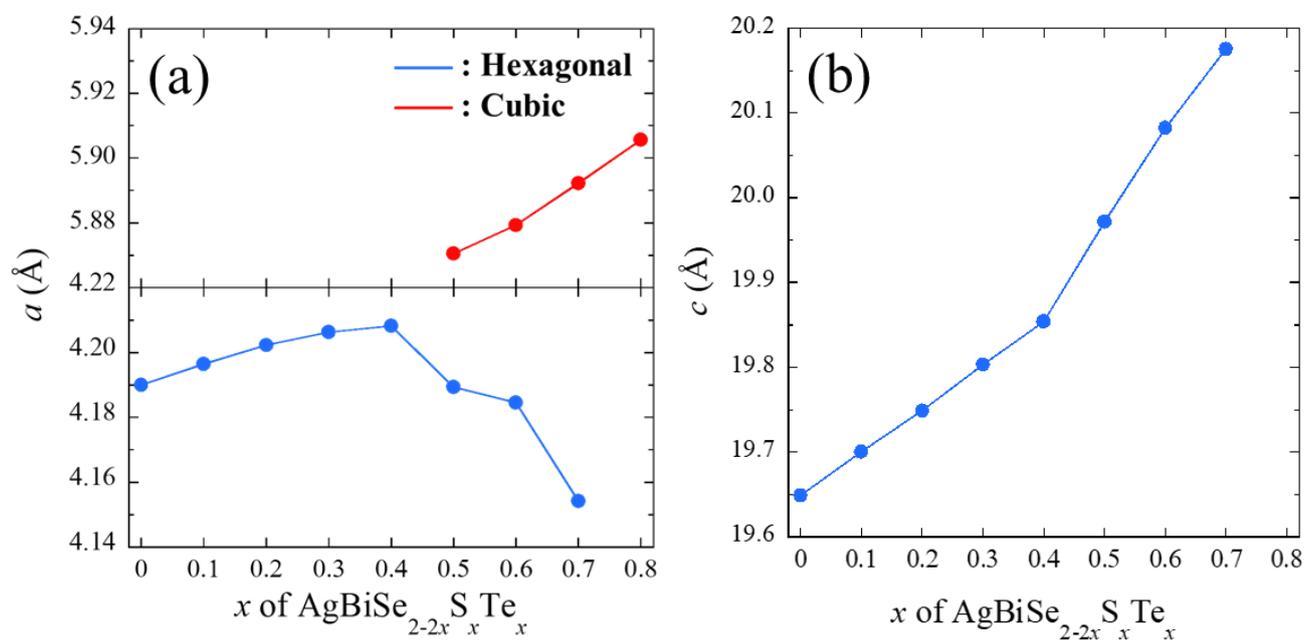

**Fig. S1** (a) $x$ amount dependence of lattice constant $a$ and (b) $c$ for hexagonal and cubic structure of AgBiSe$_{2-2x}$S$_x$Te$_x$ with $x$ = 0–0.7.

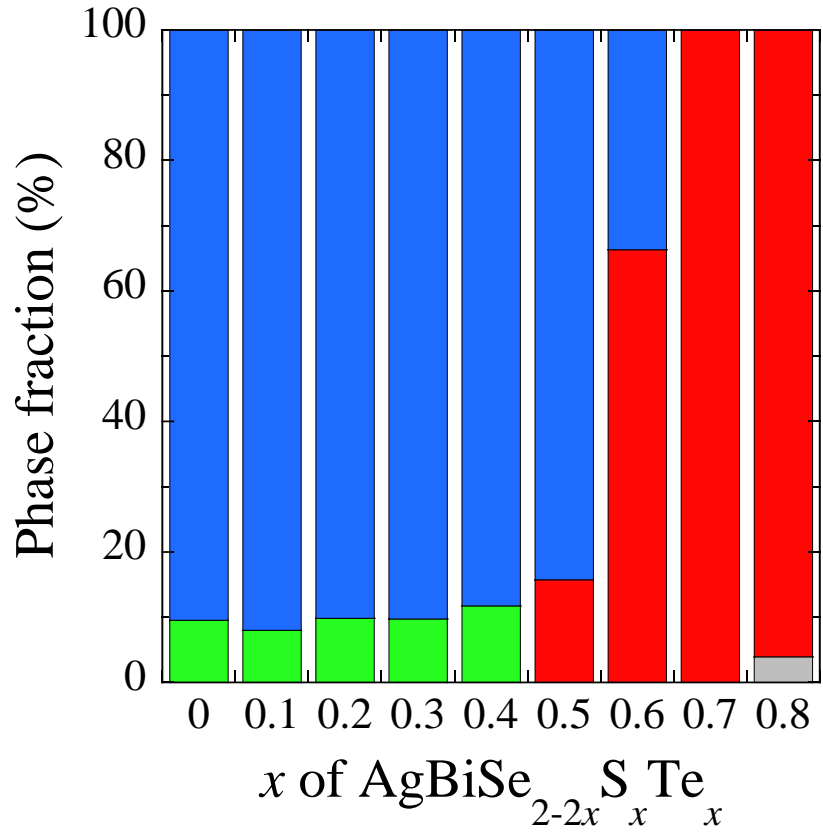

**Fig. S2** Phase fraction (%) of hexagonal (blue), rhombohedral (green), and cubic (red) structure as a function of x amount for AgBiSe$_{2-2x}$S$_x$Te$_x$ with $x$ = 0–0.8. Gray color in x = 0.8 indicates an impurity phase of Bi$_2$Ch$_3$ (Ch: S, Se, Te).

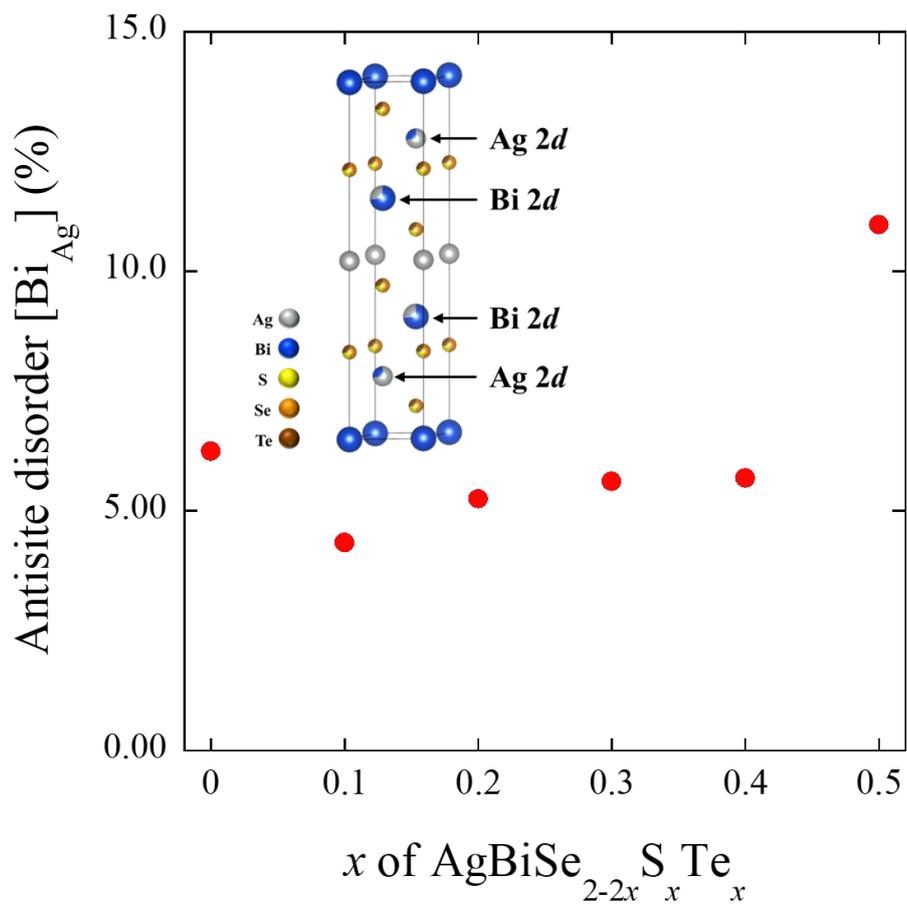

**Fig. S3** Anti-site disorder on the Bi 2d and Ag 2d crystallographic sites of hexagonal phase of AgBiSe$_{2-2x}$S$_x$Te$_x$ with $x$ = 0–0.5.

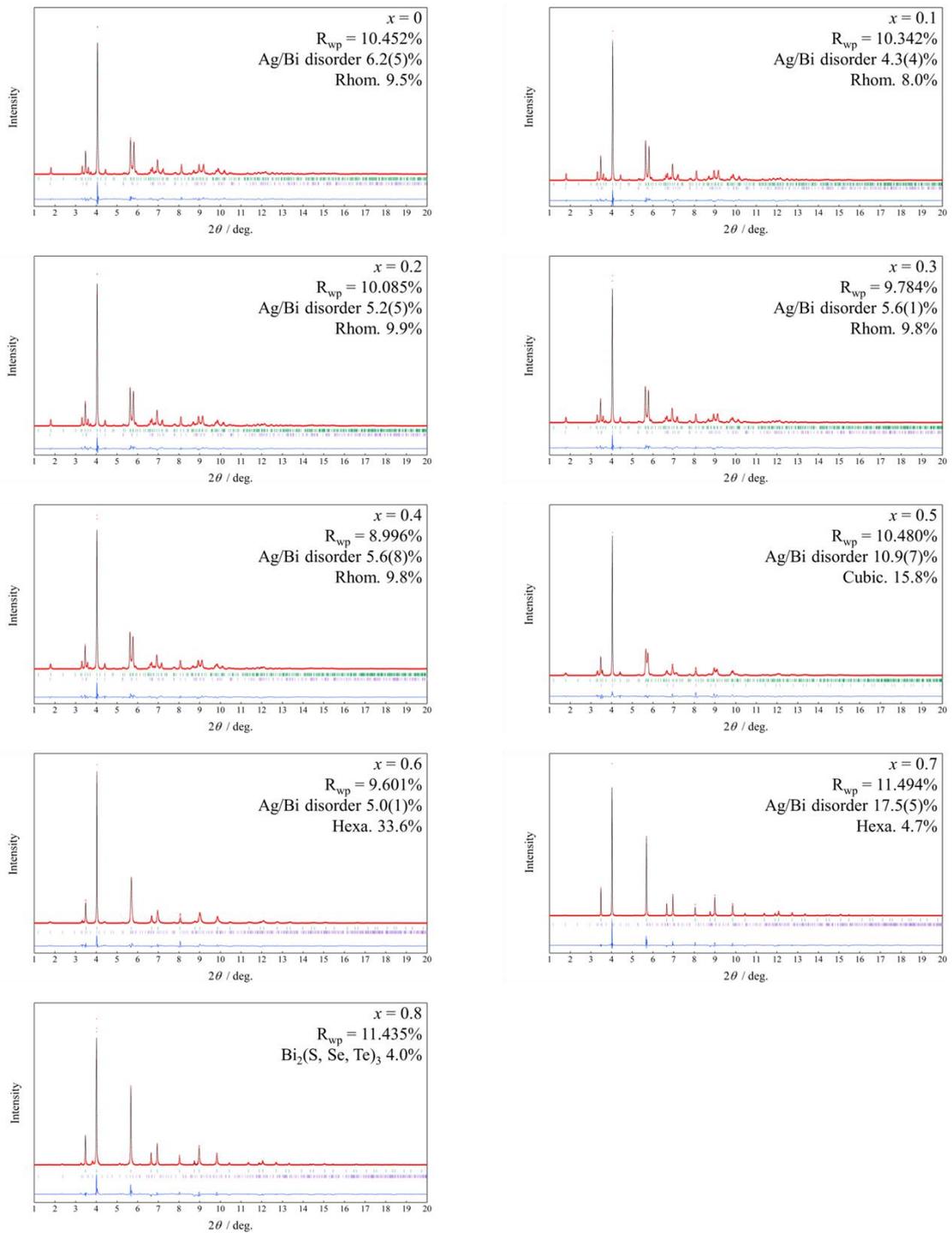

**Fig. S4** Synchrotron X-ray diffraction patterns of AgBiSe$_{2-2x}$S$_x$Te$_x$ with $x$ = 0–0.8.

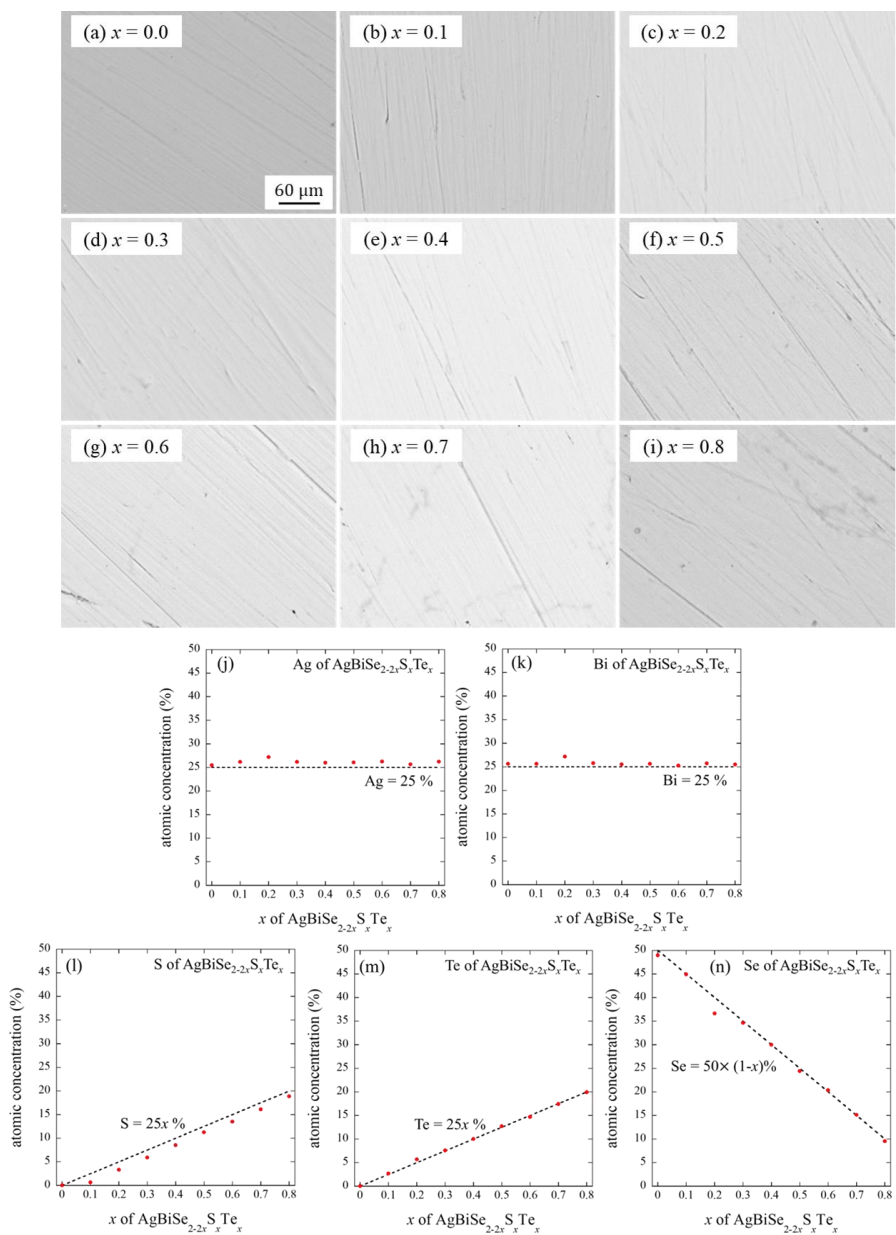

**Fig. S5** (a–i) Back scattering electron (BSE) images of polished surface of AgBiSe$_{2-2x}$S$_x$Te$_x$ with $x$ = 0.0–0.8. (j–m) Atomic concentration (%) of obtained samples by energy-dispersive X-ray spectroscopy (EDX). Note that dashed lines indicate an ideal composition.

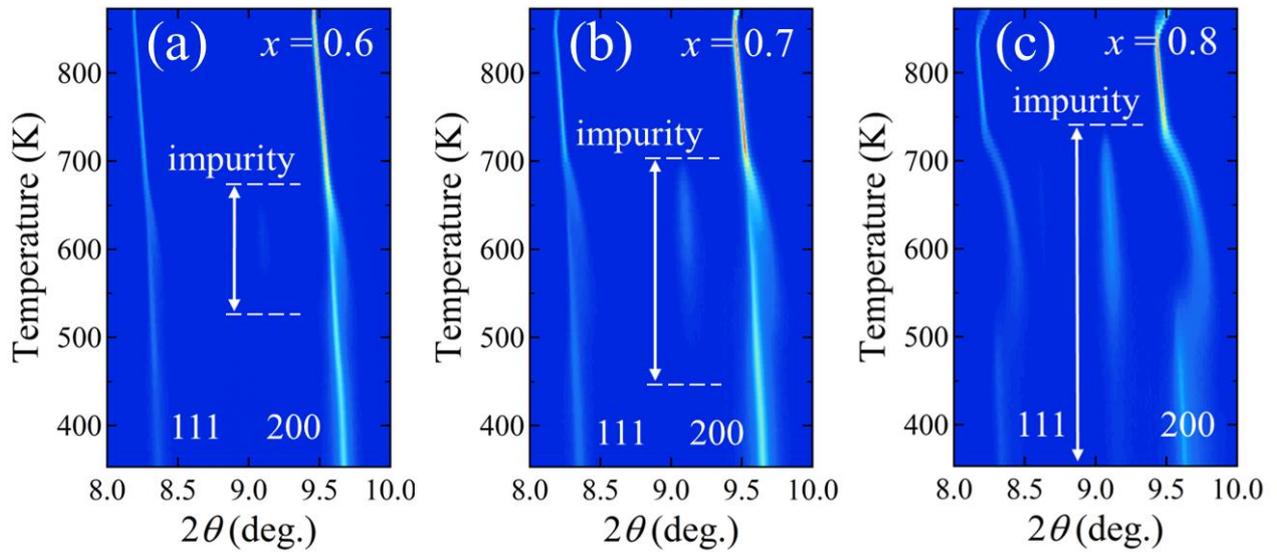

**Fig. S6** (a–c) 2D color mapping for temperature dependence synchrotron powder X-ray diffraction (SPXRD) patterns of AgBiSe$_{2-2x}$S$_x$Te$_x$ with $x$ = 0.6, 0.7, and 0.8. Note that the impurity indicates Bi$_2$(S,Se,Te)$_3$.

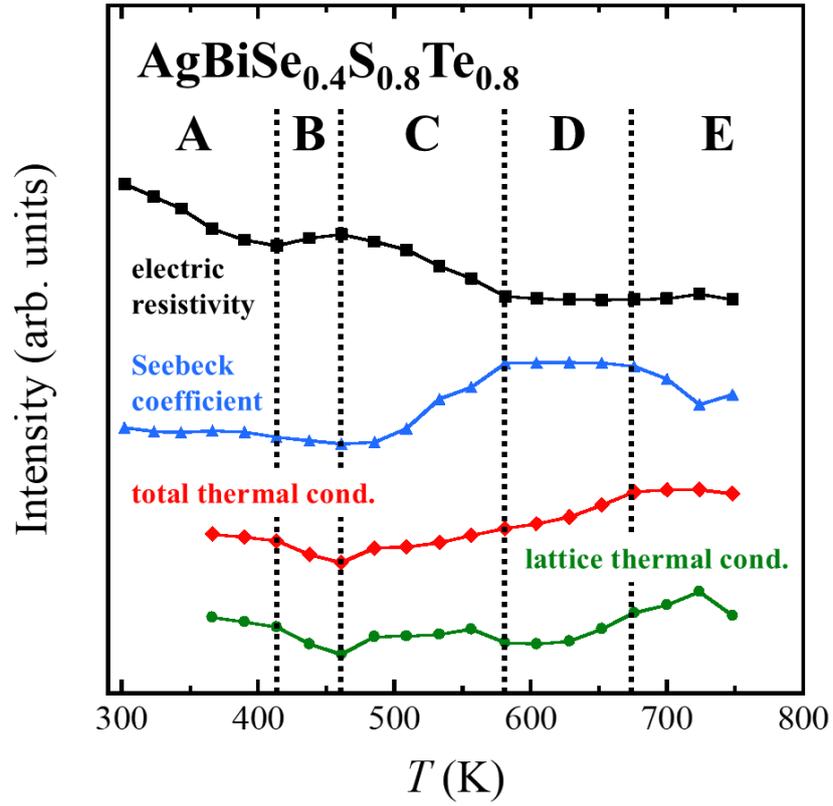

**Fig. S7** Normalization temperature dependence of electric resistivity (black square), Seebeck coefficient (blue triangle), total thermal conductivity (red rhomb), and lattice thermal conductivity (green circle) of AgBiSe$_{2-2x}$S$_x$Te$_x$ with $x$ = 0–0.8.

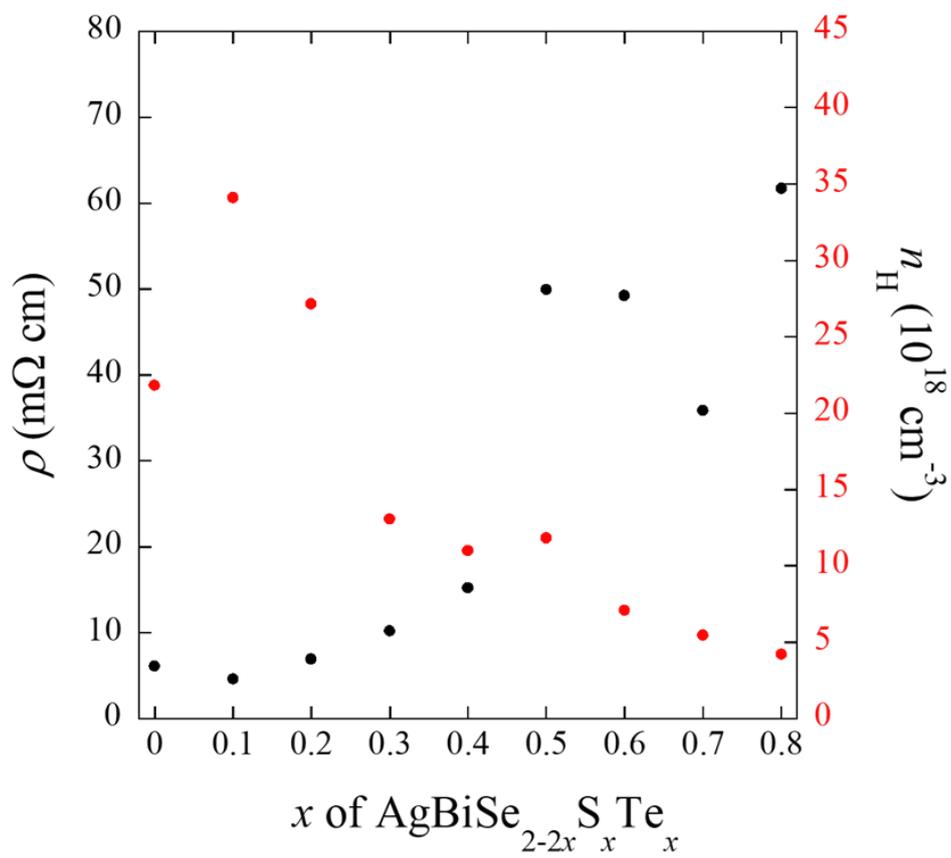

**Fig. S8** $x$ amount dependence of electrical resistivity $\rho$ (left axis) and carrier concentration $n_H$ (right axis) of AgBiSe$_{2-2x}$S$_x$Te$_x$ with $x$ = 0–0.8 at room temperature.

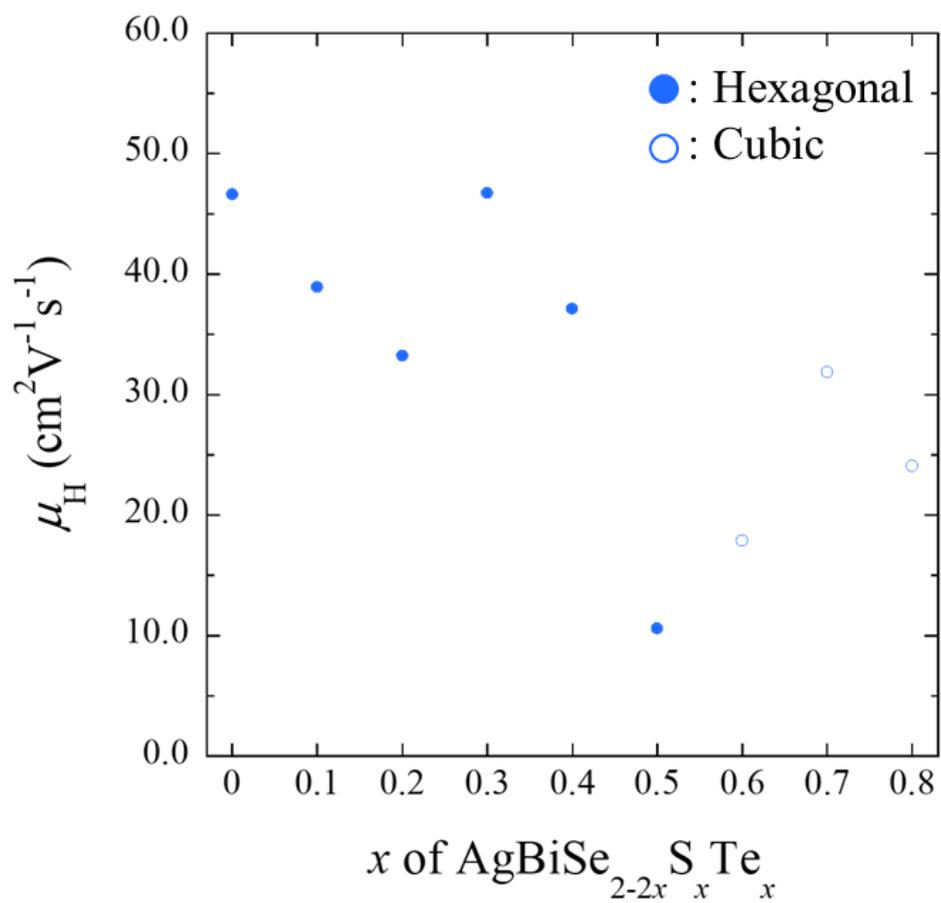

**Fig. S9** *x* amount dependence of mobility $\mu_H$ of AgBiSe$_{2-2x}$S$_x$Te$_x$ with *x* = 0–0.8 at room temperature.

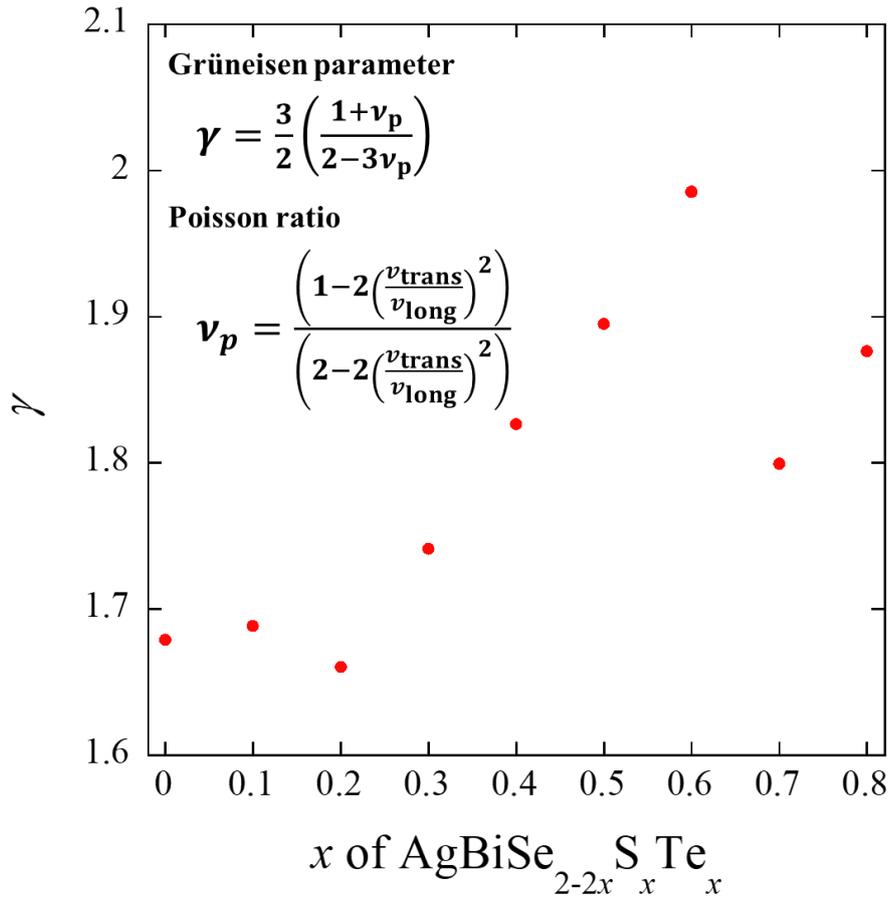

**Fig. S10** Estimated Grüneisen parameter $\gamma$ for AgBiSe$_{2-2x}$S$_x$Te$_x$ with $x = 0$–0.8.

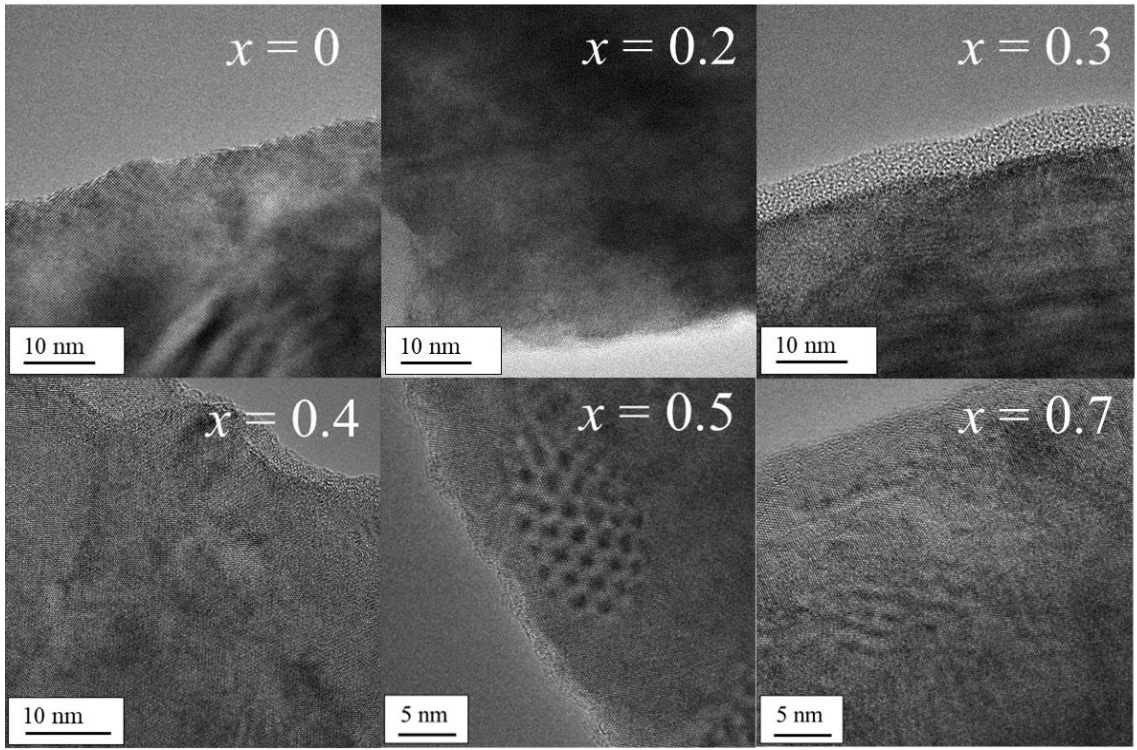

**Fig. 11** High-resolution TEM (HRTEM) observation images for AgBiSe$_{2-2x}$S$_x$Te$_x$ with $x$ = 0.0, 0.2, 0.3, 0.4, 0.5, and 0.7.

## References

[1] A. Yamashita, Y. Goto, A. Miura, C. Moriyoshi, Y. Kuroiwa, Y. Mizuguchi, n-Type thermoelectric metal chalcogenide (Ag,Pb,Bi)(S,Se,Te) designed by multi-site-type high-entropy alloying, Mater. Res. Lett. 9 (2021) 366–372. https://doi.org/10.1080/21663831.2021.1929533.

[2] Y. Luo, S. Hao, S. Cai, T.J. Slade, Z.Z. Luo, V.P. Dravid, C. Wolverton, Q. Yan, M.G. Kanatzidis, High Thermoelectric Performance in the New Cubic Semiconductor AgSnSbSe3 by High-Entropy Engineering, J. Am. Chem. Soc. 142 (2020) 15187–15198. https://doi.org/10.1021/jacs.0c07803.

[3] B. Jiang, Y. Yu, J. Cui, X. Liu, L. Xie, J. Liao, Q. Zhang, Y. Huang, S. Ning, B. Jia, B. Zhu, S. Bai, L. Chen, S.J. Pennycook, J. He, High-entropy-stabilized chalcogenides with high thermoelectric performance, Science 371 (2021) 830–834. https://doi.org/10.1126/science.abe1292.

[4] B. Jiang, W. Wang, S. Liu, Y. Wang, C. Wang, Y. Chen, L. Xie, M. Huang, J. He, High figure-of-merit and power generation in high-entropy GeTe-based thermoelectrics, Science 377 (2022) 208–213. https://doi.org/10.1126/science.abq5815.

[5] B. Jiang, Y. Yu, H. Chen, J. Cui, X. Liu, L. Xie, J. He, Entropy engineering promotes thermoelectric performance in p-type chalcogenides, Nat. Commun. 12 (2021) 3234. https://doi.org/10.1038/s41467-021-23569-z.

[6] Z. Ma, Y. Luo, W. Li, T. Xu, Y. Wei, C. Li, A.Y. Haruna, Q. Jiang, D. Zhang, J. Yang, High Thermoelectric Performance and Low Lattice Thermal Conductivity in Lattice-Distorted High-Entropy Semiconductors AgMnSn$_{1-x}$Pb$_x$SbTe$_4$, Chem. Mater. 34 (2022) 8959–8967. https://doi.org/10.1021/acs.chemmater.2c02344.

[7] J. Cai, J. Yang, G. Liu, H. Wang, F. Shi, X. Tan, Z. Ge, J. Jiang, Ultralow thermal conductivity and improved ZT of CuInTe$_2$ by high-entropy structure design, Mater. Today Phys. 18 (2021) 100394. https://doi.org/10.1016/j.mtphys.2021.100394.

[8] W. Zhang, Y. Lou, H. Dong, F. Wu, J. Tiwari, Z. Shi, T. Feng, S.T. Pantelides, B. Xu, Phase-engineered high-entropy metastable FCC Cu$_{2-y}$Ag$_y$(In$_x$Sn$_{1-x}$)Se$_2$S nanomaterials with high thermoelectric performance, Chem. Sci. 13 (2022) 10461–10471. https://doi.org/10.1039/D2SC02915D.